\documentclass[prd,twocolumn,nofootinbib,superscriptaddress,fleqn,amssymb,amsfonts,amsmath]{revtex4-2}
\usepackage{graphicx,slashed,xcolor,multirow,bbold,mathtools,sidecap,tikz,bm,ulem,enumitem,booktabs,array,adjustbox}
\usepackage[colorlinks=true,linktoc=page,linkcolor=purple,citecolor=teal,urlcolor=magenta]{hyperref}
\usepackage[compat=1.1.0]{tikz-feynman}
\graphicspath{ {plots/} }

\hypersetup{
	colorlinks=true,
	linkcolor=myviolet,
	urlcolor=myviolet,
	citecolor=myviolet
}

\definecolor{mypink2}{RGB}{219, 48, 122}
\definecolor{rosso}{cmyk}{0,1,1,0.4}
\definecolor{rossos}{cmyk}{0,1,1,0.55}
\definecolor{rossoc}{cmyk}{0,1,1,0.2}
\definecolor{blu}{cmyk}{1,1,0,0.3}
\definecolor{blus}{cmyk}{1,1,0,0.6}
\definecolor{bluc}{cmyk}{1,1,0,0.1}
\definecolor{verde}{cmyk}{0.92,0,0.59,0.25}
\definecolor{verdec}{cmyk}{0.92,0,0.59,0.15}
\definecolor{verdes}{cmyk}{0.92,0,0.59,0.4}
\definecolor{mint}{rgb}{0.24, 0.71, 0.54}
\definecolor{myviolet}{RGB}{205,105,255}

\setenumerate{label=$(\roman*)$,itemsep=1pt,parsep=1pt,topsep=1pt,partopsep=1pt}
\begin{document}
\title{\color{myviolet}Type-II see-saw at $\mu^+$$\mu^-$ collider}

\newcommand{\AddrHBNI}{
	Homi Bhabha National Institute, Training School Complex, Anushakti Nagar, Mumbai 400094, India }
\newcommand{\AddrIOP}{
	Institute of Physics, Bhubaneswar, Sachivalaya Marg, Sainik School, Bhubaneswar 751005, India}

\author{Siddharth P. Maharathy}
\email{siddharth.m@iopb.res.in}
\affiliation{\AddrIOP}
\affiliation{\AddrHBNI}

\author{Manimala Mitra}
\email{manimala@iopb.res.in}
\affiliation{\AddrIOP}
\affiliation{\AddrHBNI}

\begin{abstract}
	\noindent Doubly-charged Higgs bosons have extensively been searched at the LHC. In this work, we study the sensitivity reach of the doubly-charged scalar ($H^{\pm\pm}$) in muon collider for the well-known Type-II seesaw scenario. First, we perform a cut-based analysis to predict the discovery prospect in the muon collider operating with 3 TeV center of mass energy. In addition to this, we have also performed a multivariate analysis and compare the cut-based result with the result obtained from the multivariate analysis. We find that the cut-based analysis is more significant as compared to the multivariate analysis in the large doubly-charged scalar mass region. We predict that a doubly-charged scalar mass, $M_{H^{\pm\pm}}$, upto 1450 GeV can be probed with $5\sigma$ significance for center of mass $\sqrt{s}= 3$ TeV and integrated luminosity $\mathcal{L} = 1000\,\textrm{fb}^{-1}$.
\end{abstract}

\maketitle
\section{\label{sec:1}Introduction}
The discovery of the Higgs boson at the Large Hadron Collider~(LHC) assured the Brout-Englert-Higgs~(BEH) mechanism to be the most accurate formalism responsible for the generation of the Standard Model~(SM) fermions and gauge-bosons masses. The BEH mechanism can generate Dirac mass for neutrino by extending the SM with right-handed neutrinos, however, to explain eV scale small neutrino masses, a very tiny Yukawa ~ $10^{-12}$ is required, enhancing the fine-tuning problem of the SM to a multi-fold level. One of the most appealing mechanisms to describe tiny neutrino mass is via seesaw, where light neutrino masses are generated from higher dimensional $ d=5 $ Weinberg operator~\cite{Weinberg:1979sa}. 

The tree level realization of the Weinberg operator are Type-I~\cite{Minkowski:1977sc,Gell-Mann:1979vob,Mohapatra:1979ia,Schechter:1980gr}, Type-II~\cite{Cheng:1980qt,Schechter:1980gr,Mohapatra:1980yp} and Type-III~\cite{Foot:1988aq} seesaw mechanisms, where the  SM has been extended with a $SU(2)_L$ singlet fermion,  $SU(2)_L$ triplet scalar with hyper-charge $Y=1$ and a $SU(2)_L$ triplet fermion with hyper-charge $Y=0$ respectively. Among these, in the second variant of the seesaw mechanism, referred as the Type-II seesaw, generated light neutrino mass is proportional to the {\it vev} of the triplet scalar folded with the respective Yukawa coupling. 

Since the triplet scalar has gauge coupling, hence it can be  produced abundantly and then decays to SM particles. Depending on the triplet \textit{vev} the doubly-charged scalars can decay to same-sign di-lepton, same-sign gauge bosons and can also decay  through cascade decay. The phenomenology of the doubly-charged scalar has extensively been studied for LHC~\cite{Melfo:2011nx,Huitu:1996su,Gunion:1996pq,Chakrabarti:1998qy,Muhlleitner:2003me,Akeroyd:2005gt,Han:2007bk,delAguila:2008cj,FileviezPerez:2008jbu,Akeroyd:2009hb,Akeroyd:2010ip,Aoki:2011pz,Akeroyd:2011zza,Chiang:2012dk,Akeroyd:2012nd,Chun:2012zu,delAguila:2013mia,Chun:2013vma,Kanemura:2013vxa,Kanemura:2014goa,Kanemura:2014ipa,Kang:2014lwn,Han:2015hba,Han:2015sca,Mitra:2016wpr,Ghosh:2017pxl,Antusch:2018svb,BhupalDev:2018tox,deMelo:2019asm,Primulando:2019evb,Chun:2019hce,Padhan:2019jlc,Ashanujjaman:2021txz,Ashanujjaman:2022ofg}, $e^+$$e^-$~\cite{Nomura:2017abh,Blunier:2016peh,Crivellin:2018ahj,Agrawal:2018pci,Rahili:2019ixf,Ashanujjaman:2022tdn} collider and also for $e$$p$ machine~\cite{Dev:2019hev,Yang:2021skb,Cai:2017mow,Deppisch:2015qwa}. Several searches have been performed at the LHC by both ATLAS and CMS collaboration, and the lack of any excess from the SM signal imposes stringent bounds on the parameter space~\cite{ATLAS:2012hi,Chatrchyan:2012ya,ATLAS:2014kca,ATLAS:2014vih,CMS:2016cpz,ATLAS:2017xqs,CMS:2017fhs,CMS-PAS-HIG-16-036,ATLAS:2018ceg,ATLAS:2021jol,ATLAS:2022yzd}. The ATLAS multi lepton search~\cite{ATLAS:2017xqs} sets a limit of 770-870 GeV and 450 GeV on the mass of $H^{\pm \pm}$ considering $B(H^{\pm\pm} \rightarrow l^\pm l^\pm) = 100\%$ and $B(H^{\pm\pm} \rightarrow l^\pm l^\pm) = 10\%$ respectively. Taking into account the $W^\pm W^\pm$ or $W^\pm Z$ decay mode of the charged scalars, the ATLAS search~\cite{ATLAS:2021jol} excludes $H^{\pm \pm}$ in the range 200- 350 GeV and 230 GeV, for pair and associated production modes. For 3000 $\textrm{fb}^{-1}$ luminosity, with the aforementioned decay mode of the charged scalars,  expected reach of the doubly-charged scalar increases up to 640 GeV \cite{Ashanujjaman:2021txz}. The most stringent bound from CMS is from the CMS multilepton search~\cite{CMS-PAS-HIG-16-036}, limiting $H^{\pm \pm}$ mass to 535-820 GeV. With the assumption of equal branching ratios to each possible leptonic final states, $Br(H^{\pm\pm}\rightarrow e^{\pm}e^{\pm})=Br(H^{\pm\pm}\rightarrow e^{\pm}\mu^{\pm})=Br(H^{\pm\pm}\rightarrow e^{\pm}\tau^{\pm})=Br(H^{\pm\pm}\rightarrow \mu^{\pm}\mu^{\pm})=Br(H^{\pm\pm}\rightarrow \mu^{\pm}\tau^{\pm})=Br(H^{\pm\pm}\rightarrow \tau^{\pm}\tau^{\pm})$, the ATLAS multi-lepton search~\cite{ATLAS:2022yzd} imposes a lower limit of 1080 GeV on doubly charged scalars mass.

Other than a $pp$ machine, the doubly-charged Higgs can also be searched at  future
lepton colliders: Linear or Circular electron positron collider \cite{Behnke:2013xla,CEPCStudyGroup:2018rmc,CERN_FCC_webpage,CLIC_webpage} and muon colliders \cite{Palmer:1996gs,Ankenbrandt:1999cta,Delahaye:2019omf,Long:2020wfp,AlAli:2021let}. For higher TeV scale masses of the doubly charged Higgs for which  cross-section at the LHC is $\sim \mathcal{O}$(fb) or even lower, and to probe hadronic final states, lepton collider can be most useful, 
as depending upon the centre of mass energy,  the cross-section for doubly charged Higgs production is typically  large until the kinematic threashold. Additionally, a lepton collider can be more effective because of its cleaner signal. In case of muon collider the collision is free from any pile-up events. An additional benefit is, contrary to the circular $e^{+}e^{-}$ colliders,  a muon collider suppresses the loss of energy due to synchrotron radiation because of the heavier mass of muon, thereby making both high energy as well as high luminosity achievable. In this work we choose a particular configuration of muon collider: center-of-mass energy ($\sqrt{s}$) = 3 TeV and integrated luminosity ($\mathcal{L}$) = 1000 $\textrm{fb}^{-1}$ and explore the sensitivity  reach for the doubly charged scalar. Some recent works on BSM particle searches at  muon collider can be found in \cite{Costantini:2020stv,Bandyopadhyay:2020mnp,Han:2020uid}. 

The paper is arranged in the following way. In Sec.\,[\ref{sec:2}] we discuss the model description of Type-II seesaw. We discuss the collider analysis that includes both the cut based as well as multivariate analysis in Sec.\,[\ref{sec:3}]. Finally in Sec.\,[\ref{sec:4}]
we summarize our outcomes.

\section{\label{sec:2}Model}
The Type-II seesaw model has an extended scalar sector, where in addition to the SM scalar doublet, $\Phi = (\Phi^+\, \, \Phi^0)^T$,  a $SU(2)$ triplet scalar with hyper-charge Y=1 is also present. 
\begin{eqnarray}
 \Delta=\begin{pmatrix} \frac{\Delta^+}{\sqrt{2}} & \Delta^{++} \\ \Delta^0 & -\frac{\Delta^+}{\sqrt{2}} 
\end{pmatrix}
\end{eqnarray}
The neutral components of the scalars are parameterised as $\Phi^0 = (v_{\Phi} +  \phi^0  + iZ_1) /\sqrt{2}$ and $\Delta^0 = (v_{\Delta} +  \delta^0  + iZ_2) /\sqrt{2}$. The vacuum expectation values (\textit{vevs}) of the SM doublet and the BSM triplet are $v_{\Phi}$ and $v_{\Delta}$ respectively, and they satisfy the relation $v=\sqrt{v^2_{\Phi}+2v^2_{\Delta}}= 246 \, \, \rm{GeV}$. 

The kinetic Lagrangian of the SM scalar doublet $\Phi$ and the scalar triplet $\Delta$ has the following form,
\begin{eqnarray}
\mathcal{L}_{\rm{kin}}(\Phi, \Delta)&=&(D_\mu \Phi)^\dagger (D^\mu \Phi)+\rm{Tr}[(D_\mu \Delta)^\dagger (D^\mu \Delta)]~.
\end{eqnarray}
In the above, the covariant derivatives are given by, 
\begin{eqnarray}
D_\mu \Phi &=&\left(\partial_\mu+i\frac{g}{2}\tau^a W_\mu^a+i\frac{g'}{2}B_\mu\right)\Phi~, \nonumber \\ 
D_\mu \Delta&=&\partial_\mu \Delta+i\frac{g}{2}[\tau^a W_\mu^a,\Delta]+ig'B_\mu\Delta~. 
\end{eqnarray}

The scalar potential of the model is,
\begin{eqnarray}
V(\Phi,\Delta)&=&-m_\Phi^2\Phi^\dagger\Phi+\tilde{M}^2_{\Delta}\rm{Tr}(\Delta^\dagger\Delta) 
\nonumber \\ &+&\left(\mu \Phi^Ti\tau_2\Delta^\dagger \Phi+\rm{h.c.}\right)
+\frac{\lambda}{4}(\Phi^\dagger\Phi)^2 \nonumber \\ &+&\lambda_1(\Phi^\dagger\Phi)\rm{Tr}(\Delta^\dagger\Delta)+\lambda_2\left[\rm{Tr}(\Delta^\dagger\Delta)\right]^2 
\nonumber\\
&+&\lambda_3\rm{Tr}[(\Delta^\dagger\Delta)^2]
+\lambda_4\Phi^\dagger\Delta\Delta^\dagger\Phi~.
\label{eqn:scalpt}
\end{eqnarray}

After the symmetry breaking the charged scalars and neutral scalars mix  resulting several physical mass eigen states. The masses can be obtained by 
diagonalising with rotation matrices, $R^{\pm}$, $R^0$ and $R^0_A$, respectively: 

\begin{eqnarray}
R^{\pm}&=&
\left(
\begin{array}{cc}
c_{\beta{_\pm}}&-s_{\beta{_\pm}} \\
s_{\beta{_\pm}}& c_{\beta{_\pm}}
\end{array}
\right),
R^0=
\left(
\begin{array}{cc}
c_{\beta_{0}} & -s_{\beta_{0}} \\
s_{\beta_{0}} & c_{\beta_{0}}
\end{array}
\right) 
\end{eqnarray} 
\begin{eqnarray}
R^0_A&=&
\left(
\begin{array}{cc}
c_{\alpha} & -s_{\alpha} \\
s_{\alpha}   & c_{\alpha}
\end{array}
\right)
\end{eqnarray}
where $c_{\beta_{\pm}}$ = $\cos \beta_{\pm}$, $c_{\beta_{0}}$ = $\cos \beta_{0}$, $c_{\alpha}$ = $\cos{\alpha}$. The mixing angles $\beta_{\pm}$, $\beta_0$ and $\alpha$ have the following forms: 
\begin{eqnarray}
\sqrt{2}\tan\beta_\pm = \tan\beta_0 = \frac{2v_\Delta}{v_\Phi},~~~\tan2\alpha  = \frac{2B}{A-C}
\end{eqnarray}
where $A = \frac{\lambda v_\Phi^2}{2}$,~~$B = -\sqrt{2}\mu v_\Phi + (\lambda_1+\lambda_4)v_\Phi v_\Delta$ and $C = \frac{\mu v_\Phi^2}{\sqrt{2}v_\Delta}+2(\lambda_2+\lambda_3)v_\Delta^2$.
In the above $R^{\pm}$ represents the rotation matrix between charged scalar eigenstates, $R^0, R^0_A$ represent the rotation matrix between the CP-even and CP-odd neutral scalar states, respectively. 

In addition to the three Goldstone bosons $G^\pm$ and $G^0$ which give masses to the gauge bosons, 
there are seven physical mass eigen states $H^{\pm\pm}$, $H^\pm$, $A$, $H$ and $h$. The gauge basic and mass basic for these scalar states are related as,  
\begin{eqnarray}
\left(
\begin{array}{c}
\Phi^\pm\\
\Delta^\pm
\end{array}\right)=
R^{\pm}
\left(
\begin{array}{c}
G^\pm\\
H^\pm
\end{array}\right) &,&
\left(
\begin{array}{c}
Z_1^0\\
Z_2^0
\end{array}\right)=
R^0_A
\left(
\begin{array}{c}
G^0\\
A
\end{array}\right)\nonumber\\
\left(
\begin{array}{c}
\phi^0\\
\delta^0
\end{array}\right)&=&
R^0
\left(
\begin{array}{c}
h\\
H
\end{array}\right).
\end{eqnarray}

A detailed discussion on the mass spectrum of the physical scalars of the model has been presented in \cite{Arhrib:2011uy}. With the assumption: $v_\Delta \ll v_\Phi$, the masses of the physical scalars takes the following simplified form,
\begin{eqnarray*}
	M^2_{H^{\pm\pm}}&\simeq& \tilde{M}^2_{\Delta} - \frac{\lambda_4}{2}v_\Phi^2,~~~	M^2_{H^{\pm}}\simeq \tilde{M}^2_{\Delta} - \frac{\lambda_4}{4}v_\Phi^2,\\
	M^2_{h}&\simeq& 2 v_\Phi^2 \lambda,~~~~
	M^2_{H}\approx M^2_{H} \simeq \tilde{M}^2_{\Delta}
\end{eqnarray*}
where $ \tilde{M}^2_{\Delta} \equiv \frac{\mu v_\Phi^2}{\sqrt{2}v_\Delta}$ and the mass-squared differences are obtained as,
\begin{eqnarray}
	M^2_{H^{\pm}}-M^2_{H^{\pm\pm}}\approx M^2_{H,A}-M^2_{H^{\pm}}\approx \frac{\lambda_4}{4}v_\Phi^2
\end{eqnarray}
Considering the sign of $\lambda_4$, three possible mass spectrum of the physical scalars can be realised,
\begin{enumerate}
	\item Degenerate scenario ($\lambda_4\!=\!0$):\! $M^2_{H^{\pm\pm}}\!=\!M^2_{H^{\pm}}\!=\!M^2_{H,A}$
	\item Normal scenario ($\lambda_4\!>\!0$): $M^2_{H^{\pm\pm}}> M^2_{H^{\pm}}>M^2_{H,A}$
	\item Inverted scenario ($\lambda_4\!<\!0$): $M^2_{H^{\pm\pm}}< M^2_{H^{\pm}}<M^2_{H,A}$
\end{enumerate}

The Yukawa interaction of the model responsible for neutrino mass is,
\begin{eqnarray}
	-\mathcal{L} = Y_{ij}L_i^TCi\sigma_2\Delta L_j + h.c.
	\label{Eq:neutrino_mass}
\end{eqnarray}
where $Y$ is a $3\times3$ complex symmetric matrix and $L = (\nu_L,l_L)^T$ is the left-handed SM lepton doublet. After the triplet scalar, $\Delta$, acquires the \textit{vev} ($v_\Delta$), a tiny neutrino mass can be obtained from Yukawa term in Eq.\,\ref{Eq:neutrino_mass} as,
\begin{eqnarray}
	m_\nu = \sqrt{2}\,Y v_\Delta
\end{eqnarray}
The above $3\times3$ matrix, $m_\nu$, can further be diagonalised by unitary transformation by Pontecorvo-Maki-Nakagawa-Sakata (PMNS) matrix $U_{PMNS}$ as $U^T_{PMNS}m_\nu U_{PMNS} = m^d_\nu = diag(m_1,m_2,m_3)$, where $m_1,m_2,m_3$ are the three mass eigen values. The matrix $U_{PMNS}$ is parameterized by three mixing angles ($\theta_{12}, \theta_{23}, \theta_{13}$), one Dirac phase ($\delta$) and two Majorana phases ($\phi,\phi'$).

Various constraints on the model parameter space are the following,
\begin{itemize}
	\item \textit{Constraints from electroweak precision data (EWPD):}
	The $\rho$-parameter ($m_W^2/m_Z^2\cos^2_{\theta_W}$) of the model in terms of the doublet and triplet \textit{vev} is defined as \cite{Padhan:2019jlc,Ashanujjaman:2021txz} ,
	$$\rho = (v_\Phi^2 + 2v_\Delta^2)/(v_\Phi^2 + 4v_\Delta^2)$$
	the present \textit{EWPD} \cite{ParticleDataGroup:2020ssz} sets the value of $\rho$ parameter as, $\rho = 1.00038(20)$ which is 1.9$\sigma$ away from SM tree-level value ($\rho = 1$) and this leads to an upper bound of $\mathcal{O}(1)$ GeV on $ v_\Delta $.
	\item \textit{Constraints from oblique parameters:}
	The mass splitting, $\Delta m$, between the different physical scalars affects the \textit{EWPD} observable, named as oblique parameters ($S$, $T$ and $U$ parameters). These parameters constraints the mass splitting as $\Delta m < 40$ GeV \cite{Melfo:2011nx,Chun:2012zu}.
	\item \textit{Constraints from lepton flavour violation(LFV):}
	From the Yukawa interaction shown in Eq.\,\ref{Eq:neutrino_mass}, \textit{LFV} decays like $\mu\rightarrow e\gamma$ at loop-level and $\mu\rightarrow 3e$ at tree-level can be possible. The branching fractions can be calculated as \cite{Kakizaki:2003jk,Dinh:2012bp},
	$$\hspace{0.5cm}BR(\mu\rightarrow e\gamma) = \frac{\alpha|(Y^\dagger Y)_{e\mu}|^2}{192\pi G_F^2}\left(\frac{1}{m_{H^\pm}^2} + \frac{8}{m_{H^{\pm\pm}}^2}\right)$$
	$$BR(\mu\rightarrow 3e) = \frac{|(Y^\dagger_{ee} Y_{\mu e})|^2}{4 G_F^2 m_{H^{\pm\pm}}^4}$$
	\vspace{0.1cm}
	
	where $\alpha$ is the electromagnetic fine-structure constantand $G_F$ is the Fermi constant. The upper bounds on the above processes, $4.2 \times 10^{-13}$ for $\mu\rightarrow e\gamma$ \cite{MEG:2016leq} and $1.0 \times 10^{-12}$ for $\mu\rightarrow 3e$ \cite{SINDRUM:1987nra} limit the lower value of the triplet \textit{vev}($v_\Delta$), which can be expressed as \cite{Ashanujjaman:2021txz},
	$$\hspace{0.8cm} v_\Delta \gtrsim 0.78-1.5(0.69-2.3)\, \text{GeV} \times 10^{-9}\times\frac{1~\text{TeV}}{M_{H^{\pm\pm}}}$$
	for NH(IH).
	\item \textit{Constraints from Colliders:}
	For $\Delta m = 0$ with large(small) $v_\Delta$, doubly-charged scalar mass below 420(955) GeV has already been excluded \cite{Ashanujjaman:2021txz}. The limit extends upto 1115 GeV for $\Delta m < 0$ and moderate $v_\Delta$. However for moderate $v_\Delta$ with  $\Delta m > 0$, doubly-charged scalar as light as 200 GeV are still allowed by the LHC results.
\end{itemize}

\begin{widetext}
		
\begin{figure}[htb!] 
\centering
\includegraphics[scale=0.49]{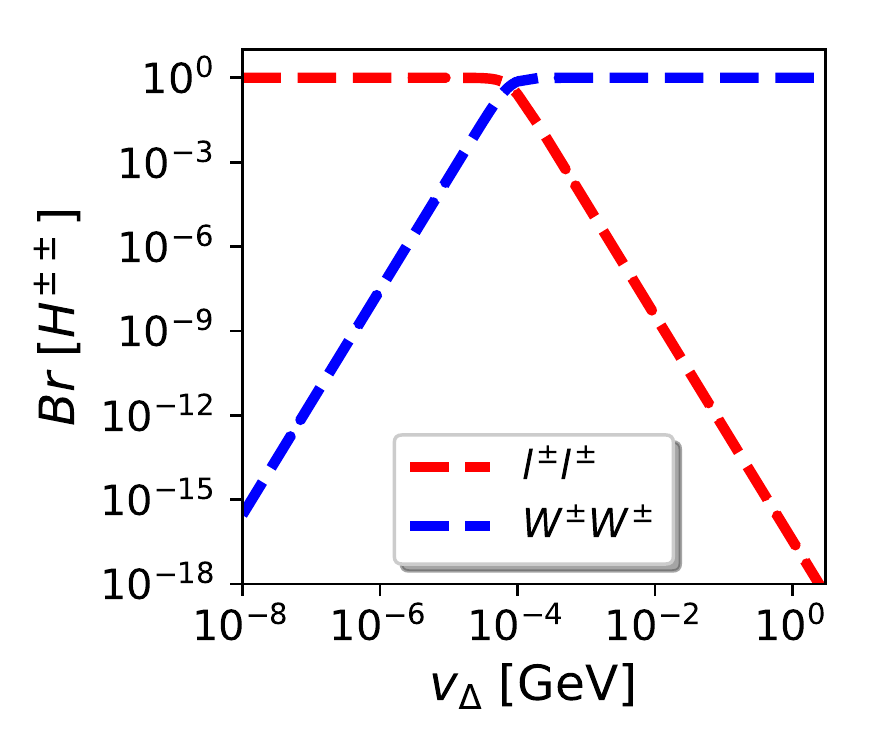}
\includegraphics[scale=0.49]{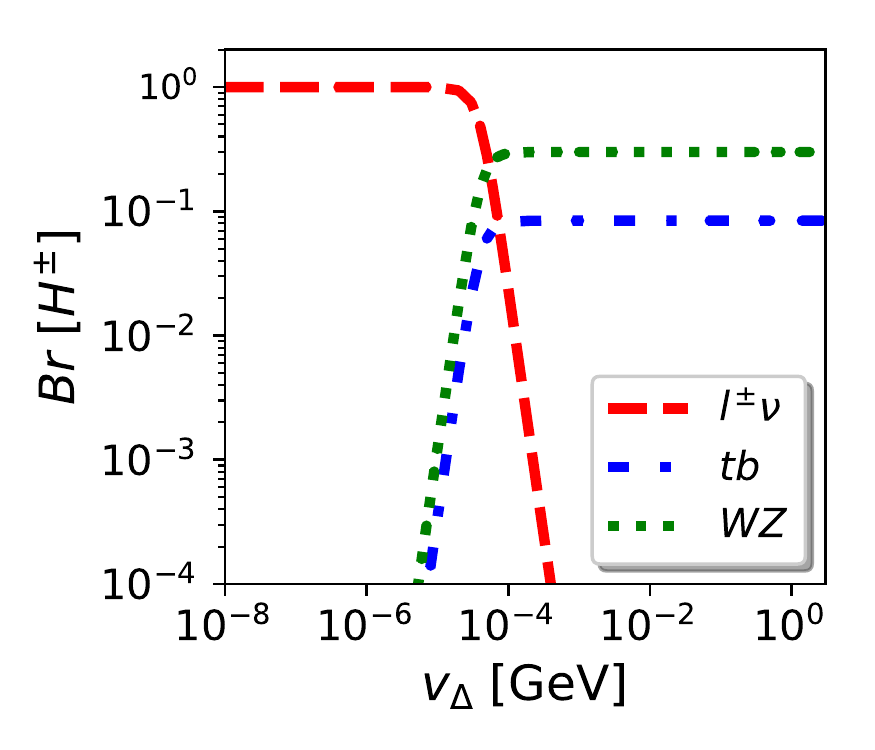}
\includegraphics[scale=0.49]{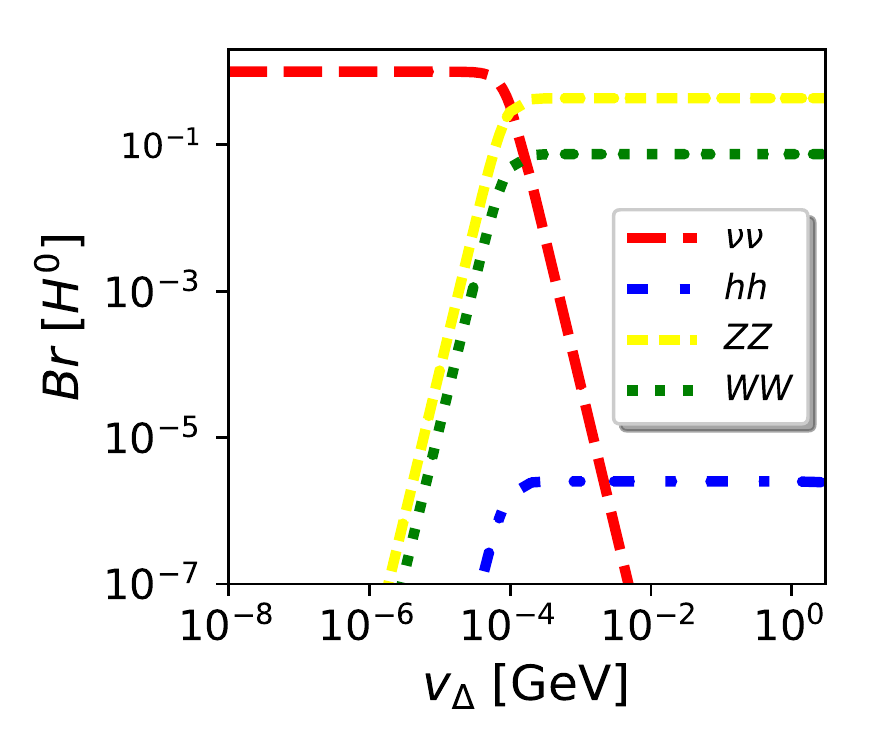}		
\includegraphics[scale=0.49]{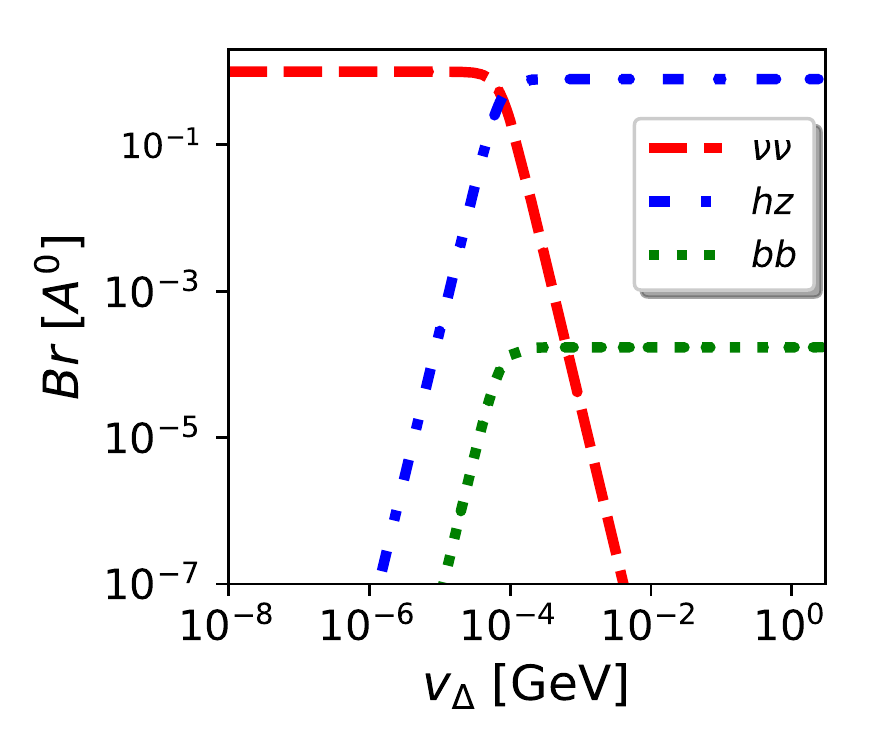}
\caption{\label{Fig:br_of_bsmscalars} Branching ratios of doubly-charged scalar: $H^{\pm\pm}$ (left panel), singly-charged scalar: $H^{\pm}$ (2nd panel), CP-even scalar: $H$ (3rd panel) and CP-odd scalar: $A$ (right panel) versus triplet \textit{vev} ($v_\Delta$) for an illustrative benchmark point $M_{H^{\pm\pm}}$ = $M_{H^{\pm\pm}}$ = $M_{H}$ = $M_{A}$ = 1000 GeV.}
\end{figure}

\end{widetext}

\subsection{Decay Widths and Branching Ratios}
In this section, we discuss the different decay modes of the  scalars: $H^{\pm\pm}$, $H^\pm$, $A$ and $H$. We are mainly interested on the degenerate mass spectrum of these scalars and hence we have not given much attention to the other mass spectrums (Normal and Inverted scenario). The decay width of the scalars are extensively studied and can be found in \cite{FileviezPerez:2008jbu,Aoki:2011pz,Rizzo:1980gz,Keung:1984hn,CMS:2019lwfF}. In Fig.\,\ref{Fig:br_of_bsmscalars}, we have shown the branching fractions of all the aforementioned scalars having a degenerate mass of 1000 GeV. From the figure it is quite evident that for $v_\Delta < 10^{-4}$ GeV, all the scalars dominantly decay into leptonic final states ($l^\pm l^\pm$, $l^\pm \nu$ and $\nu\nu$). However, for $v_\Delta > 10^{-4}$ GeV the hadronic decay modes ($t\bar{b},t\bar{t}~\text{and}~b\bar{b} $) along with the di-boson decay modes ($W W$, $W Z$, $h Z$ and $h h$ ) starts to dominate over the leptonic decay mode.

When the non-degenerate scenario is realized, \textit{viz} $\Delta m \neq 0$, cascade decay modes turn out to be important. This scenario is beyond the scope of our work, however the interested readers can look for this scenario in \cite{Ashanujjaman:2021txz}.
 
\section{\label{sec:3}Collider Analysis}
 
\begin{figure}[h]
 	\includegraphics[height=6cm,width=7cm]{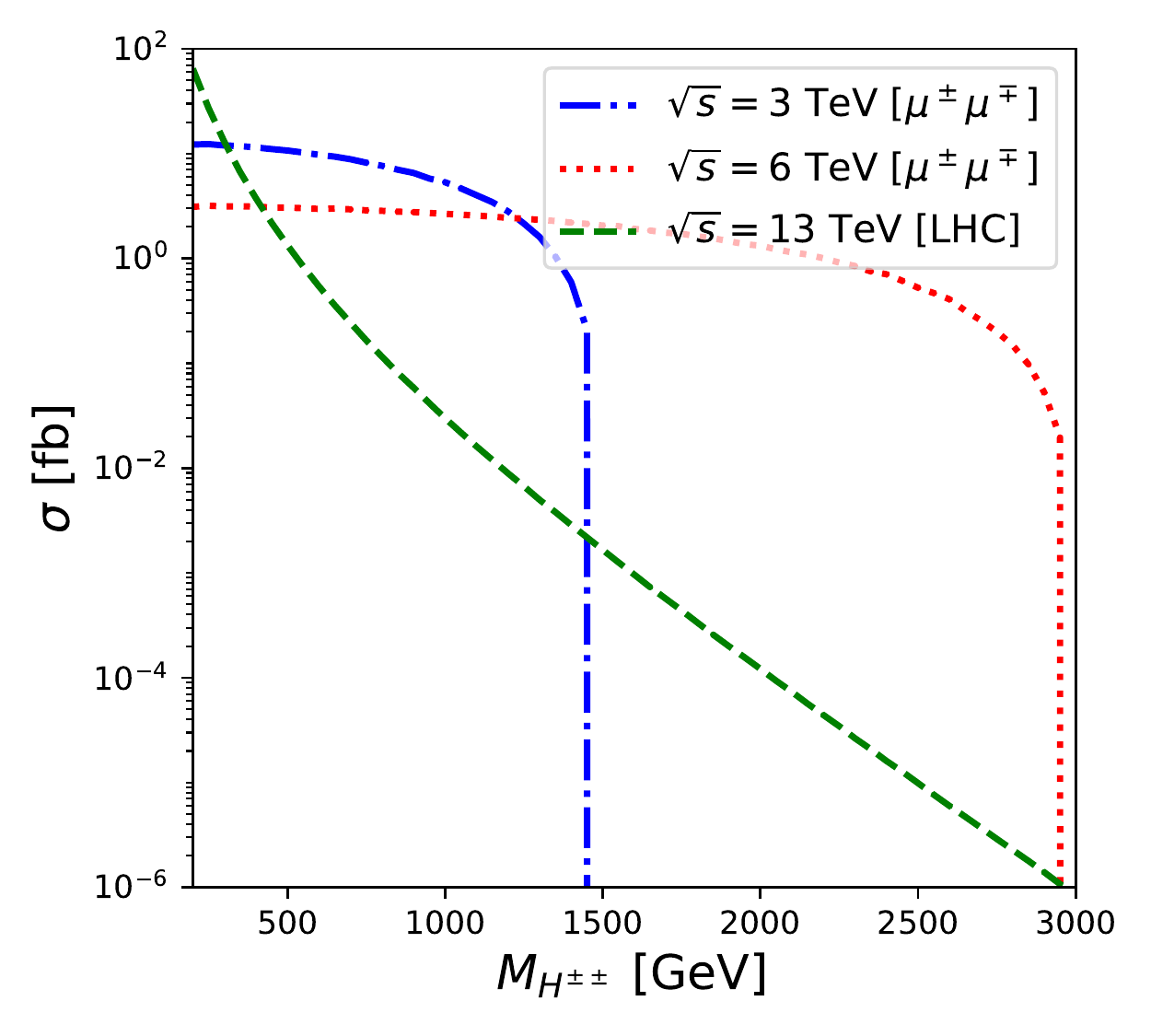}
	\caption{\label{Fig:pair_prod_crx} Production cross-section of doubly-charged scalars ($H^{\pm\pm}$) as a function of $M_{H^{\pm\pm}}$ for  3, and  6  TeV $\mu^+ \mu^-$ collider along with 13 TeV LHC. }
 	
 \end{figure}

 In this section, we discuss  the potential strength of $\mu^+$$\mu^-$ collider in probing doubly-charged scalars and exploring its sensitivity reach mainly in the high mass regime. A center of mass energy of $\sqrt{s}$ = 3 TeV is being considered for our collider analysis. The overall analysis is focused mainly in the large triplet \textit{vev} region, $v_\Delta$ $>$ $10^{-4}$, where the gauge boson mode of the doubly-charged scalars is dominant. At muon collider the doubly-charged scalars, $H^{\pm\pm}$, is produced by photon and $Z$ mediated Drell–Yan (DY) processes, shown in Fig.\,\ref{Fig:Signal_diagram}. The DY pair production cross section of the doubly-charged scalars has been shown in Fig.\,\ref{Fig:pair_prod_crx} for two different configuration of muon colliders, $\sqrt{s}$ = 3, and 6 TeV along with the production cross section at 13 TeV LHC. Being an s-channel process the production cross section decreases with the increase in center-of-mass energy. A sharp fall in the production cross section can be seen around the kinematic threshold of each configurations of muon collider i.e. around $M_{H^{\pm\pm}} \sim \frac{\sqrt{s}}{2}$. 
 As can be seen from the figure that  after $\mathcal{O}(400)$ GeV mass range, clearly the cross-section in the muon collider is substantially large.  In the following analysis we mainly focus into the multi jet final state resulting from the decay of  $H^{\pm\pm}$, as all hadronic final state in a leptonic collider can provide a  better sensitivity reach.  
  
\begin{figure}[h]
	\centering
	\includegraphics[height=5.5cm,width=6.5cm]{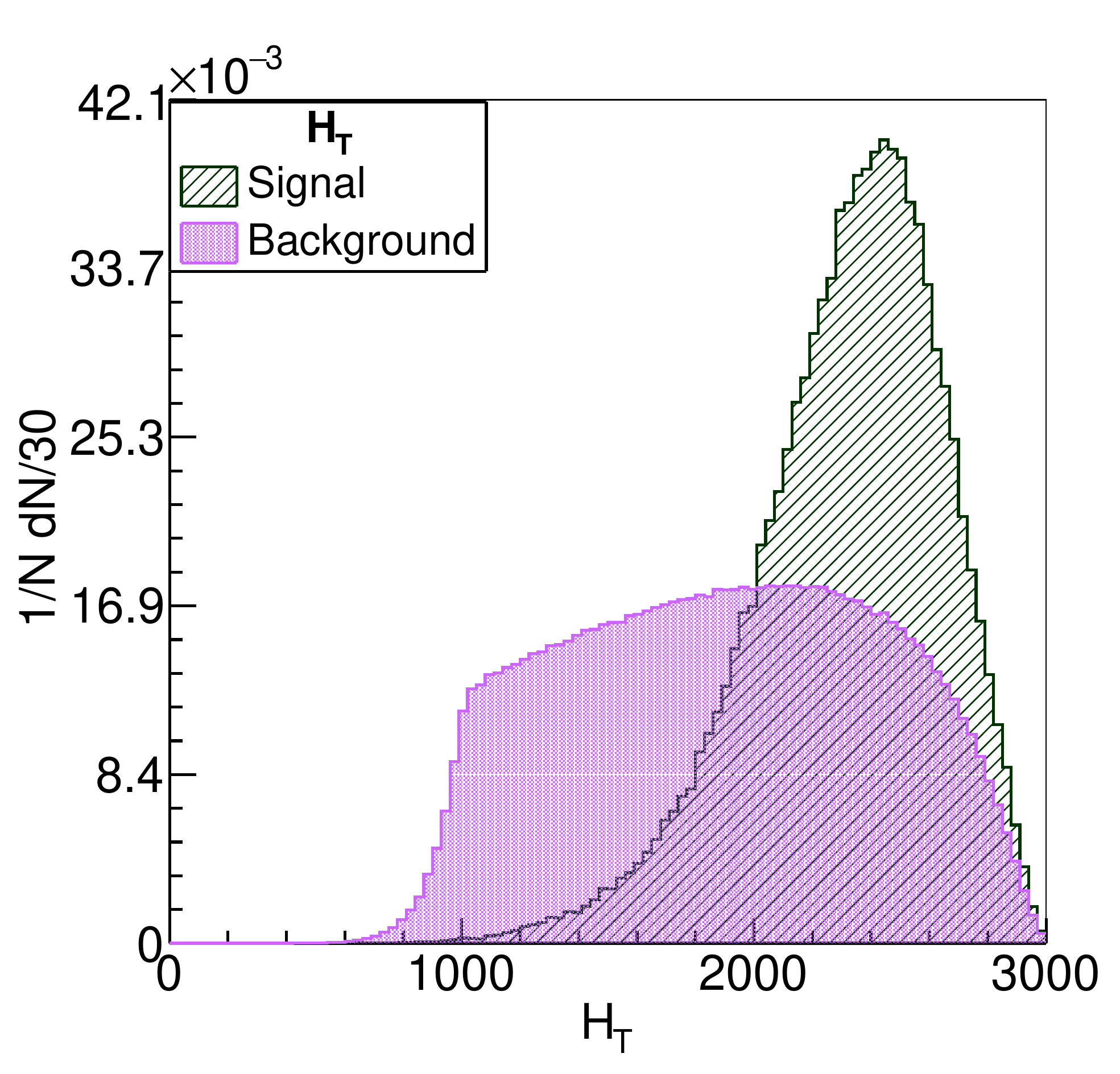}
	\caption{\label{Fig:HT_distribution}Normalized $H_T$ distribution for signal (benchmark point $M_{H^{\pm\pm}}$ = 1 TeV) and background at 3 TeV $\mu^+$$\mu^-$  collider.}
	
\end{figure}

\subsection{Multijet signature}

\begin{figure}[h]
	\centering
	\includegraphics[height=5cm,width=7cm]{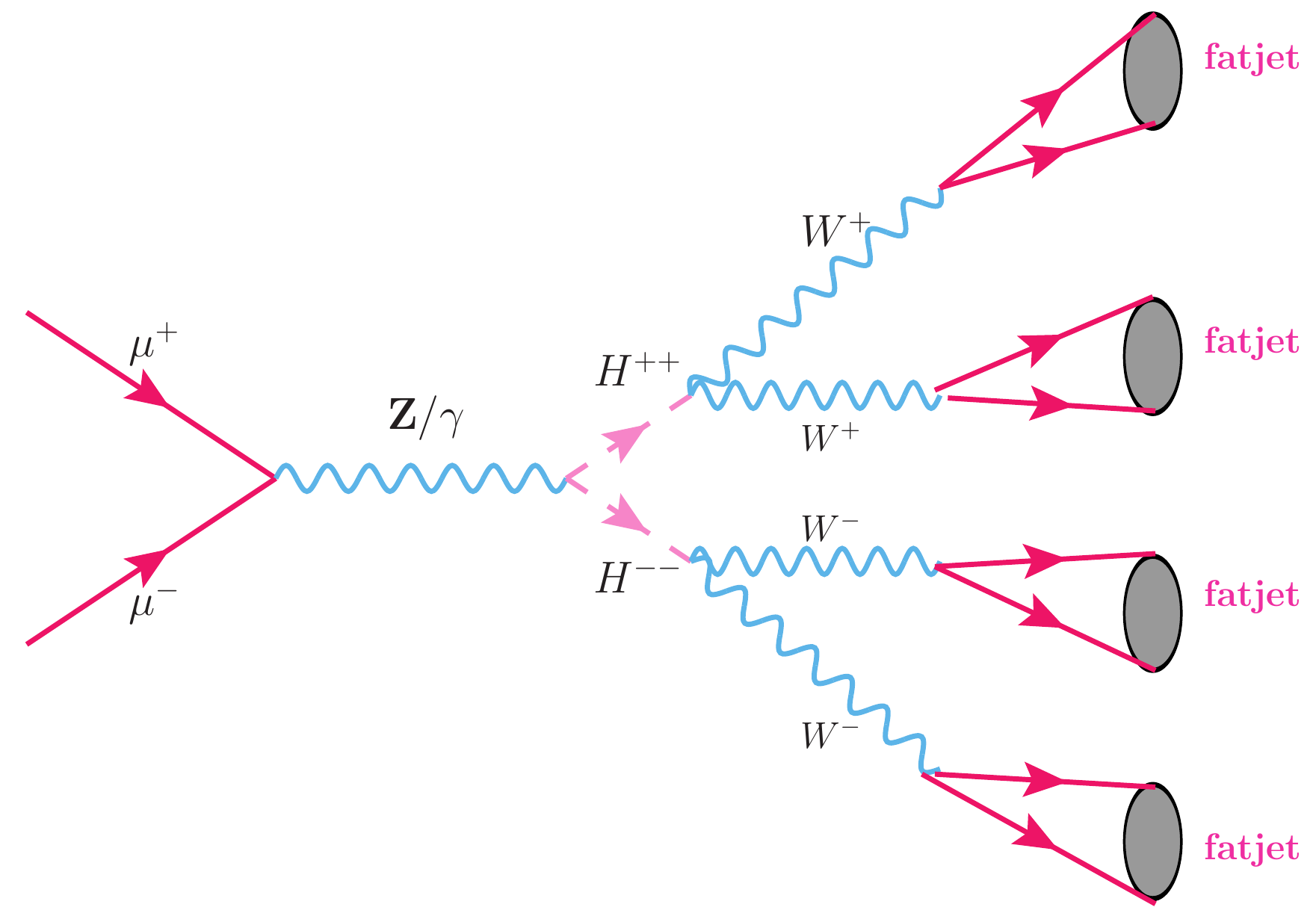}
	\caption{\label{Fig:Signal_diagram}The Feynman diagram for pair production of doubly-charged scalars ($H^{\pm\pm}$) via Drell-Yan process and their consecutive decays leading  to multijet final state. }	
\end{figure}

The analysis in the paper aim to probe the high $v_\Delta$ regime of the parameter space, the region dominated by the gauge production of $H^{\pm\pm}$ and its subsequent decay into multijet final states. The parameter space of our interest is the higher mass region of $H^{\pm\pm}$ where the production of $W^{\pm}$ is on-shell. In this region the jets resulted from $W^{\pm}$ are highly collimated, and fat-jets reconstruction is more favorable. Thus our signal comprises of up to 4 fat-jets as shown in the Fig.\,\ref{Fig:Signal_diagram}.  The production process for this signal is,  
\begin{eqnarray}
\mu^+ \mu^- \rightarrow H^{\pm\pm}H^{\mp\mp} \rightarrow 4W^{\pm} (W^{\pm} \rightarrow jj)
\end{eqnarray}
A significant number of SM background can mimic these  final states with multiple fat-jets. We consider the following sets of  backgrounds in our analysis,  
\begin{eqnarray}
\mu^+ \mu^- &\rightarrow& j j j j  \nonumber \\
\mu^+ \mu^- &\rightarrow& V + 2j \quad(V \rightarrow j j ) \nonumber \\
\mu^+ \mu^- &\rightarrow& V + 3j \quad(V \rightarrow j j ) \nonumber \\
\mu^+ \mu^- &\rightarrow& VV + 2j \quad(V \rightarrow j j ) \nonumber \\
\mu^+ \mu^- &\rightarrow& VVV + 2j \quad(V \rightarrow j j )
\end{eqnarray}
where V = ($W^{\pm},Z,h$).

\begin{widetext}
	
	\begin{figure}[htb!]
		\centering
		\includegraphics[scale=0.22]{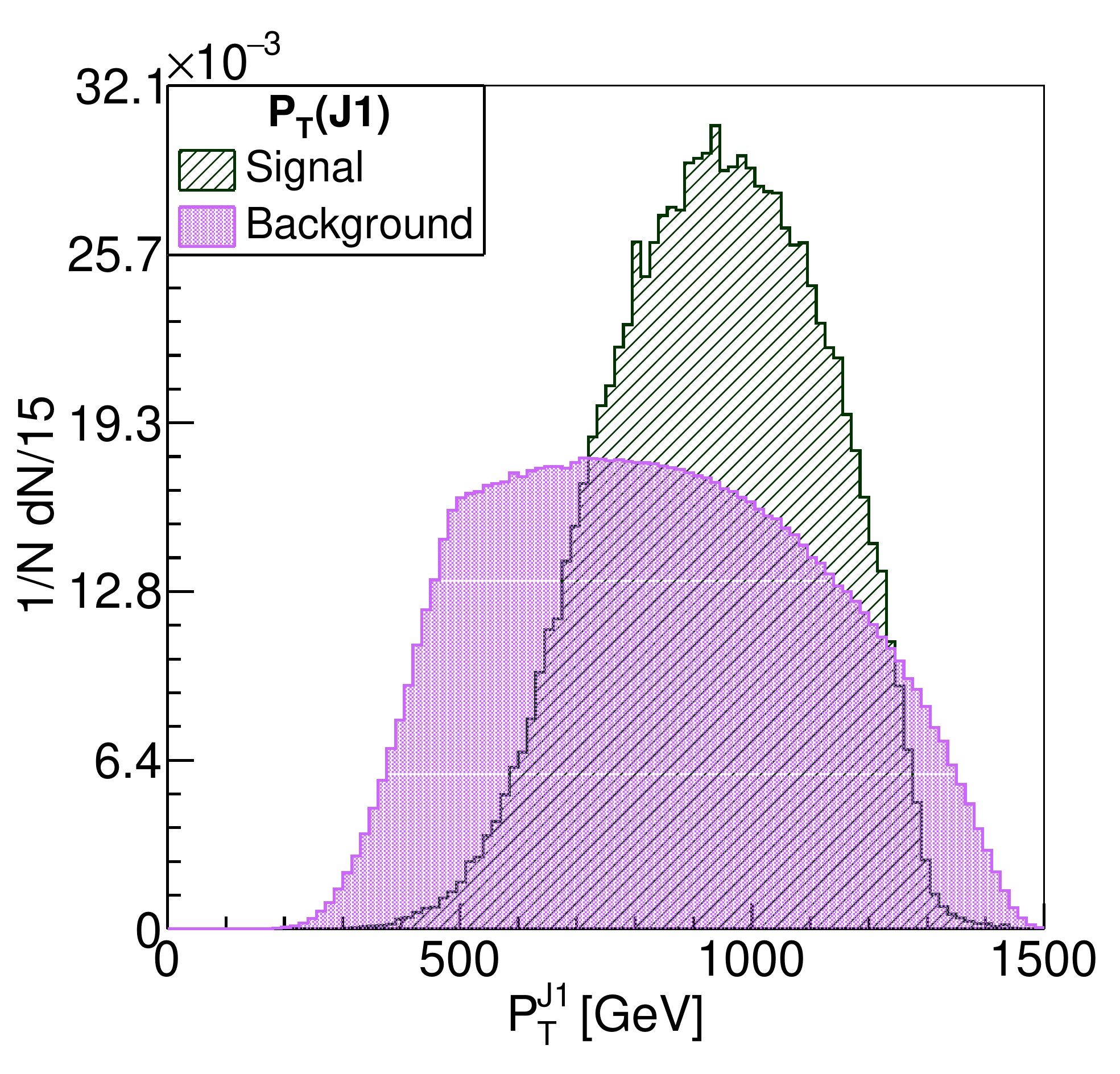}
		\includegraphics[scale=0.22]{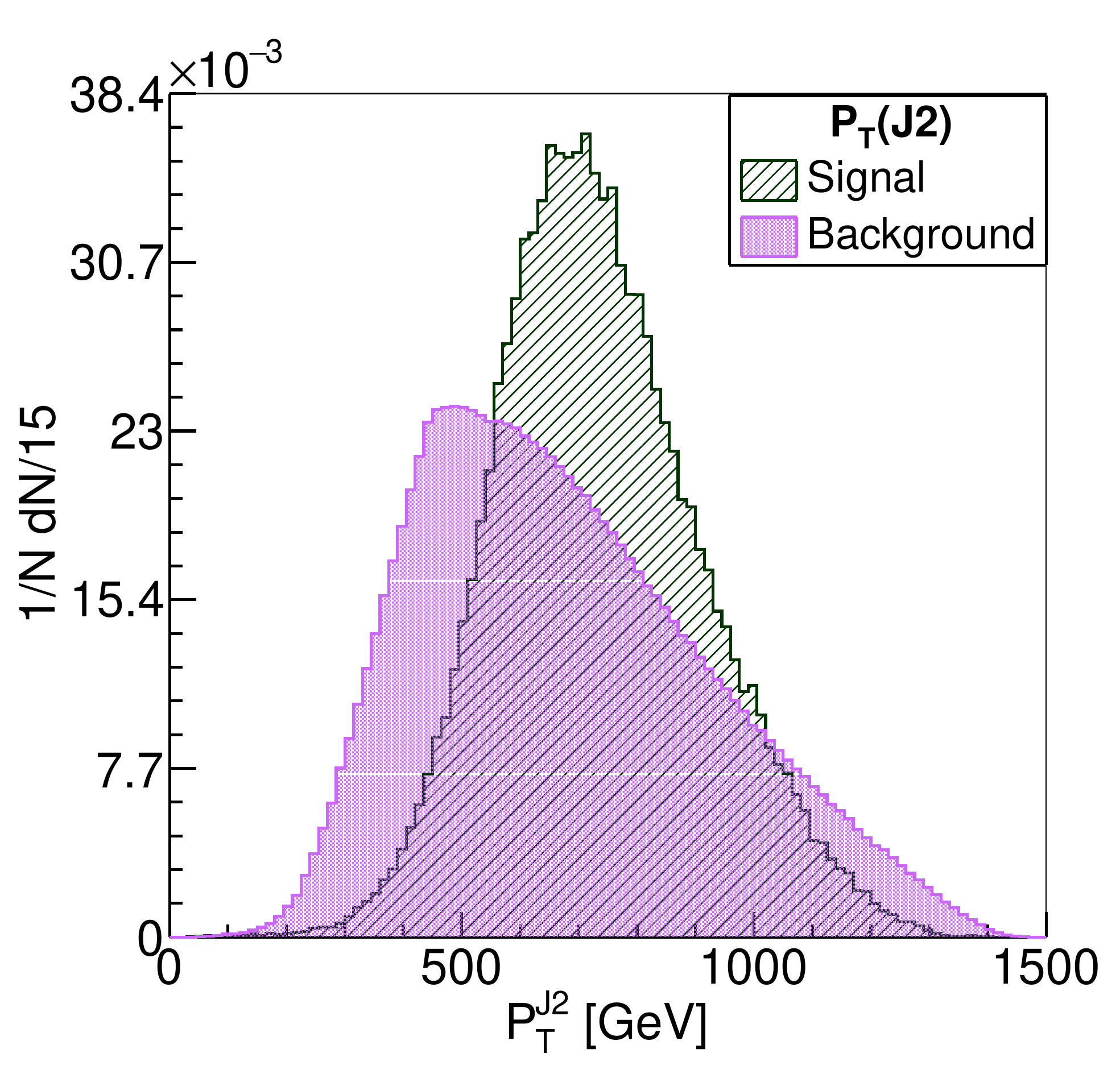}		\includegraphics[scale=0.22]{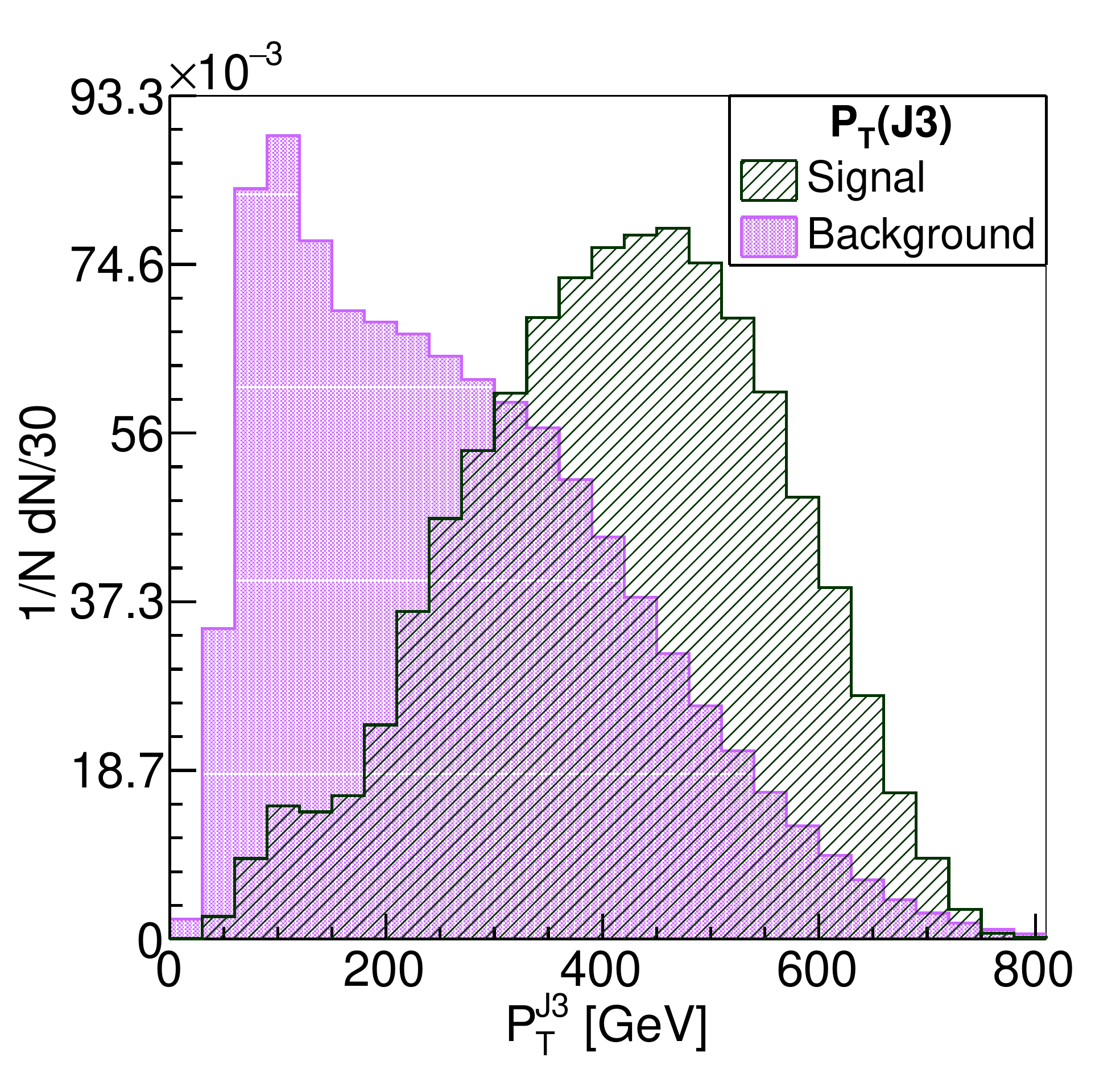}		
		\includegraphics[scale=0.22]{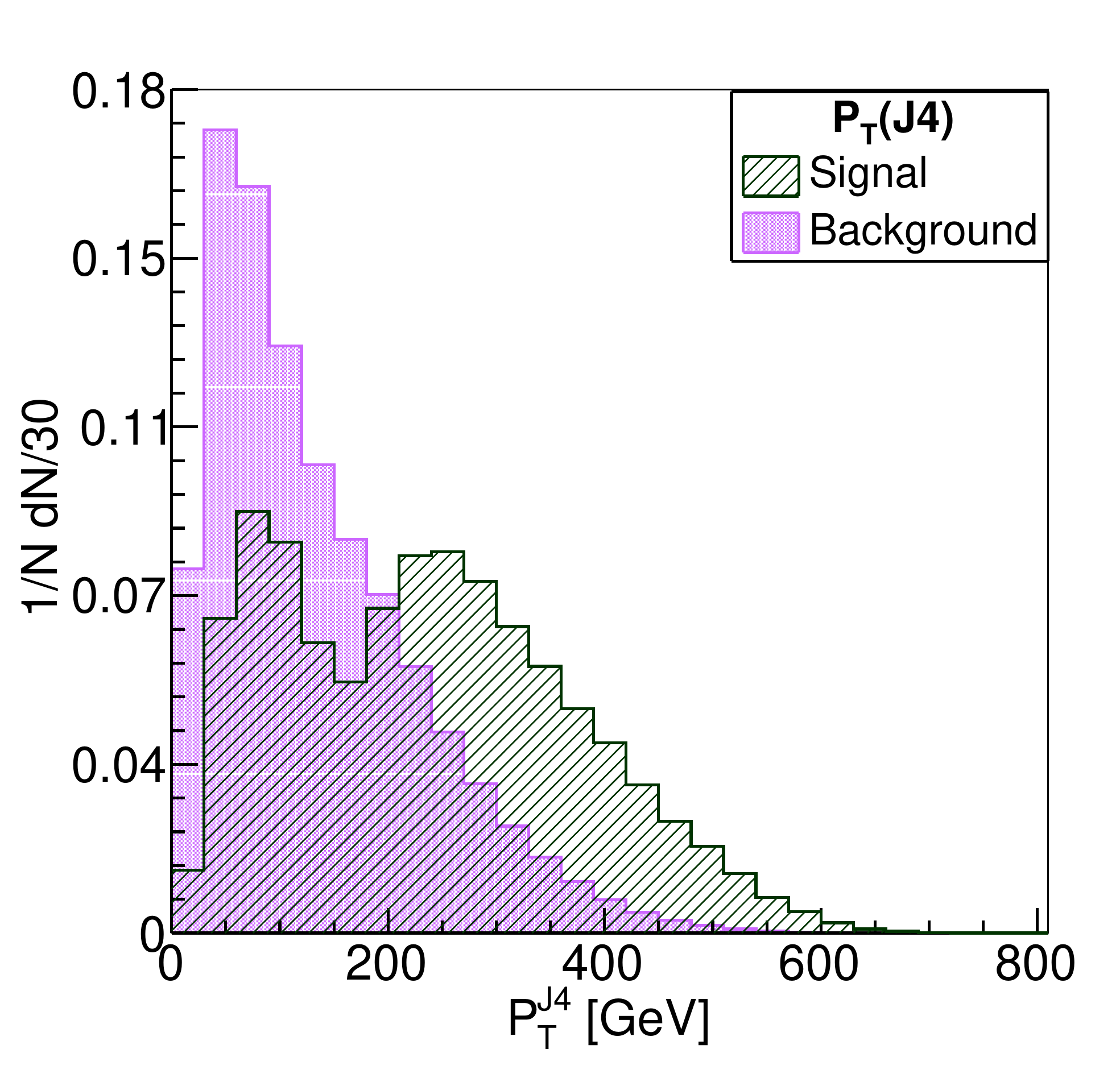}
		\caption{\label{Fig:Pt_distribution}Normalized $p_T$ distribution of leading 4-fat-jets for signal (benchmark point $M_{H^{\pm\pm}}$ = 1 TeV) at 3 TeV $\mu^+$$\mu^-$  collider.}
	\end{figure}
	
\end{widetext}

\texttt{MadGraph5aMC@NLO} \cite{Alwall:2014hca} is used to generate parton level signal and background processes. During the generation process some pre-selection cuts have been implemented to reduce the size of the background. The following enumerated points discuss about these pre-selection cuts,
\begin{enumerate}
	\item Since the signal contains at-least 4-jets (fat-jets) we imposed an increased jet-jet separation criteria at the production level, mainly for 4j background. The 4j background is generated with $\Delta R(j,j) > 0.6$ and $p_T^{j} > 60$ GeV. The  $p_T$ requirement is only on the leading 4-jets and the rest of the jets are required to have relaxed minimum  $p_T$ value i.e. $p_T \ge 20$ GeV. For all other background processes the  jet-jet separation is $\Delta R(j,j) > 0.4$ and the leading 4-jets have a minimum $p_T$ value, $p_T^{j} > 60$ GeV. The minimum $p_T$ requirement on the background processes is justified, as most of the signal is mainly populated in the high $p_T$ region, as can be seen from Fig.\,\ref{Fig:Pt_distribution}.	
	\item Fig.\,\ref{Fig:HT_distribution}  corresponds to scalar sum of transverse momenta of all visible objects, $H_T$, of both signal and backgrounds. Most of the signals are populated at high $H_T$ region as compared to the background. Hence we have also imposed a minimum $H_T$ criteria of $H_T  > 1000$ GeV on all the background processes at the generation level i.e. during the production of backgrounds at Madgraph. 

\end{enumerate}

After the parton level event generation in \texttt{MadGraph5aMC@NLO\,\cite{Alwall:2014hca}}, we then pass the generated events into \texttt{Pythia8}\,\cite{Sjostrand:2014zea} for showering and hadronization. For simulating detector effects we use \texttt{Delphes3}\,\cite{deFavereau:2013fsa}, and reconstructed jets, electrons, muons and missing energy ($E_T^{miss}$). The purpose is accomplished by using the \texttt{Delphes} \texttt{ILD} card.  We use \texttt{FastJet}\,\cite{Cacciari:2011ma} for the clustering of jets and consider Cambridge-Aachen (CA) jet clustering algorithm \,\cite{Dokshitzer:1997in} with radius parameter R = 0.8. All the jets are required to be in the pseudorapidity interval $|\eta| < 2.5$ and to have $p_T > 20$ GeV. All the leptons, both electrons and muons, have $p_T > 10$ GeV and $|\eta| < 2.5$. The missing transverse momenta ($p_T^{miss}$) is calculated from the momentum imbalance along the transverse direction for all the reconstructed objects.  As we have mainly focused on the multi-jet final state we did not bother about the isolation requirement of the leptons. 

\subsubsection{Cut based analysis} 

After reconstruction of all the physical objects (mostly jets), we than follow the cut and count method to discriminate our signal from the background. The cut flow  for two benchmark mass points, $M_H^{\pm\pm} = 1000\,\text{GeV}$ and $M_H^{\pm\pm} = 1400\, \text{GeV}$ are given in Table.\,\ref{tab:cut_flow}. From Fig.\,\ref{Fig:Pt_distribution} it can be seen that the  4th jet is quite soft. Hence imposing a hard cut on all the reconstructed 4 fat-jets,  we found it more effective to put large  $p_T$ cuts on the leading jets and a relaxed  $p_T$ cut on the sub-leading ones.

\begin{widetext}
	
	\begin{figure}[htb!]
		\centering
		\includegraphics[scale=0.35]{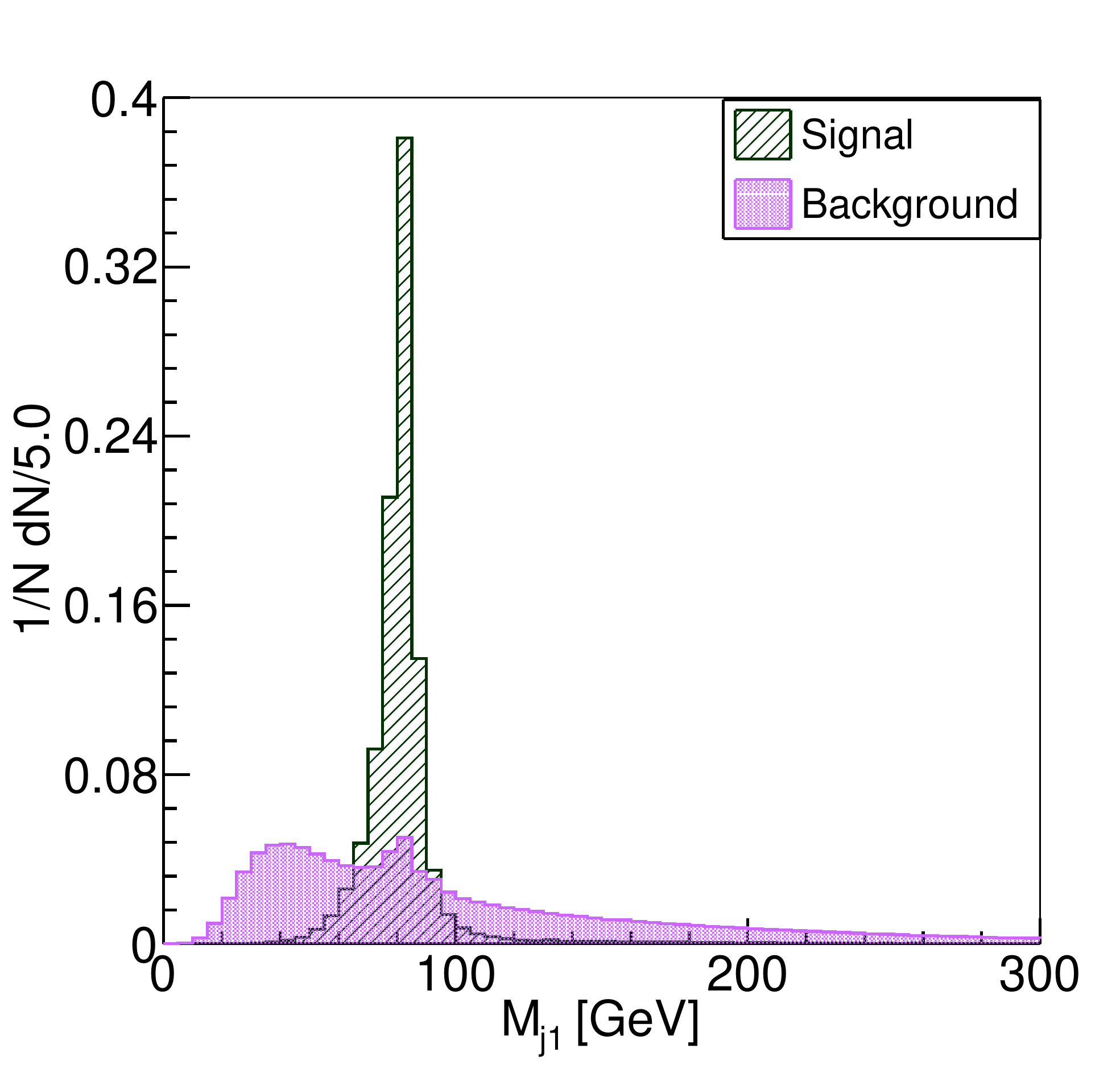}
		\includegraphics[scale=0.35]{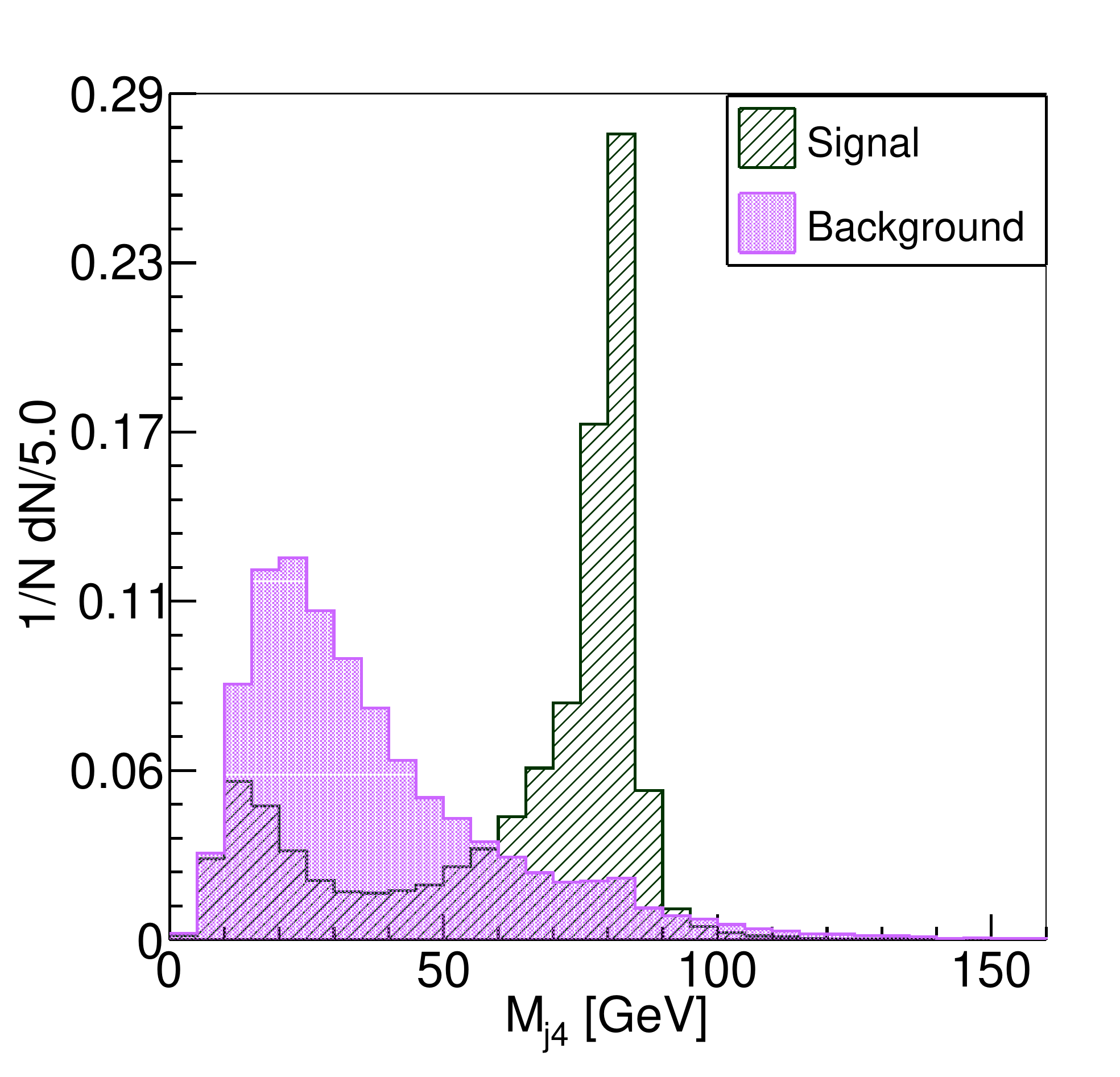}
		\caption{\label{Fig:jetinvmass_distribution}Normalized invariant mass distribution of $j_1$ and $j_4$ for signal ($M_{H^{\pm\pm}}$ = 1 TeV) and background at 3 TeV center of mass energy of $\mu^+$$\mu^-$  collider.}
	\end{figure}
	
\end{widetext}

\begin{widetext}
	
	\begin{figure}[htb!]
		\includegraphics[scale=0.35]{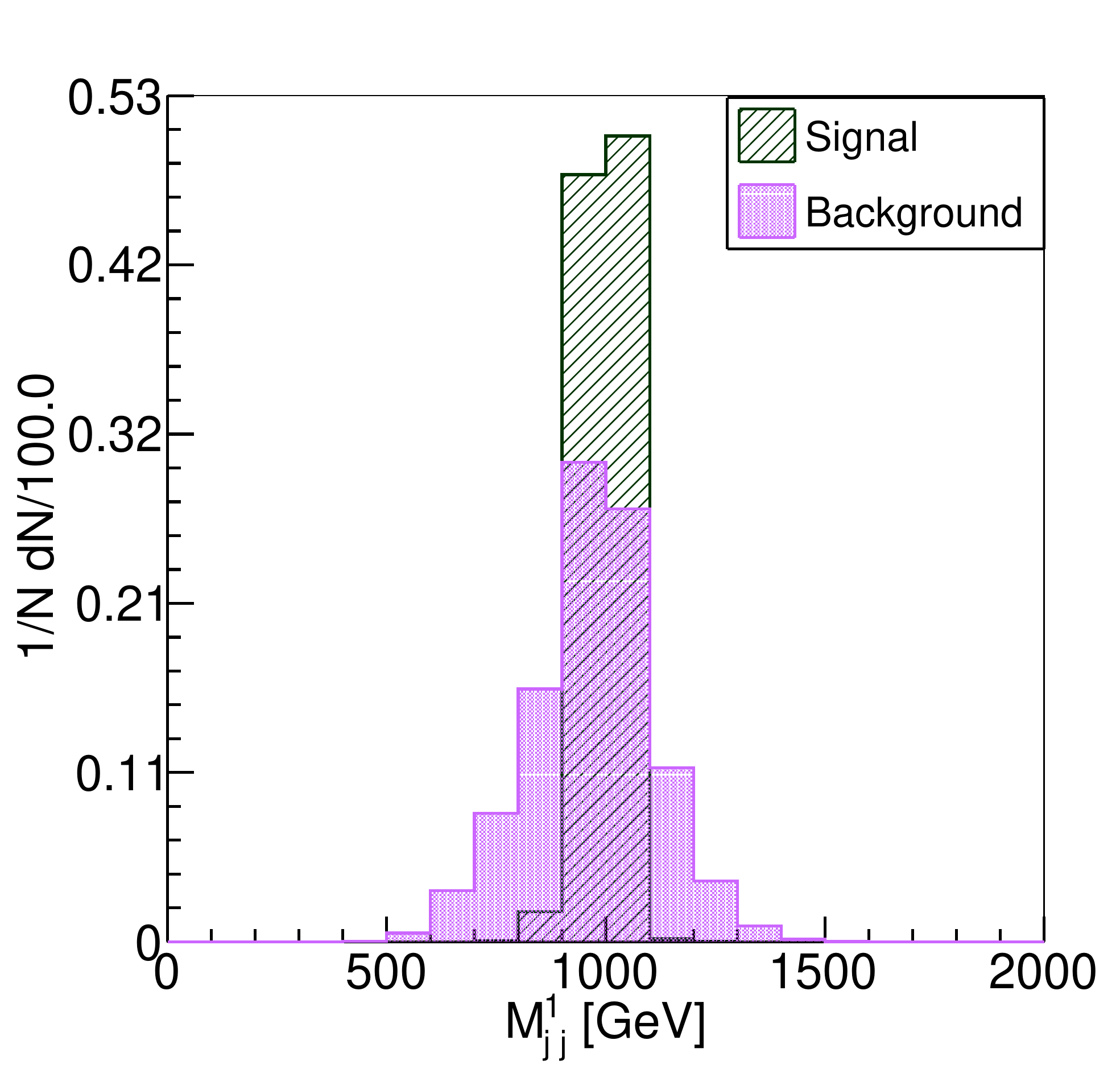}
		\includegraphics[scale=0.35]{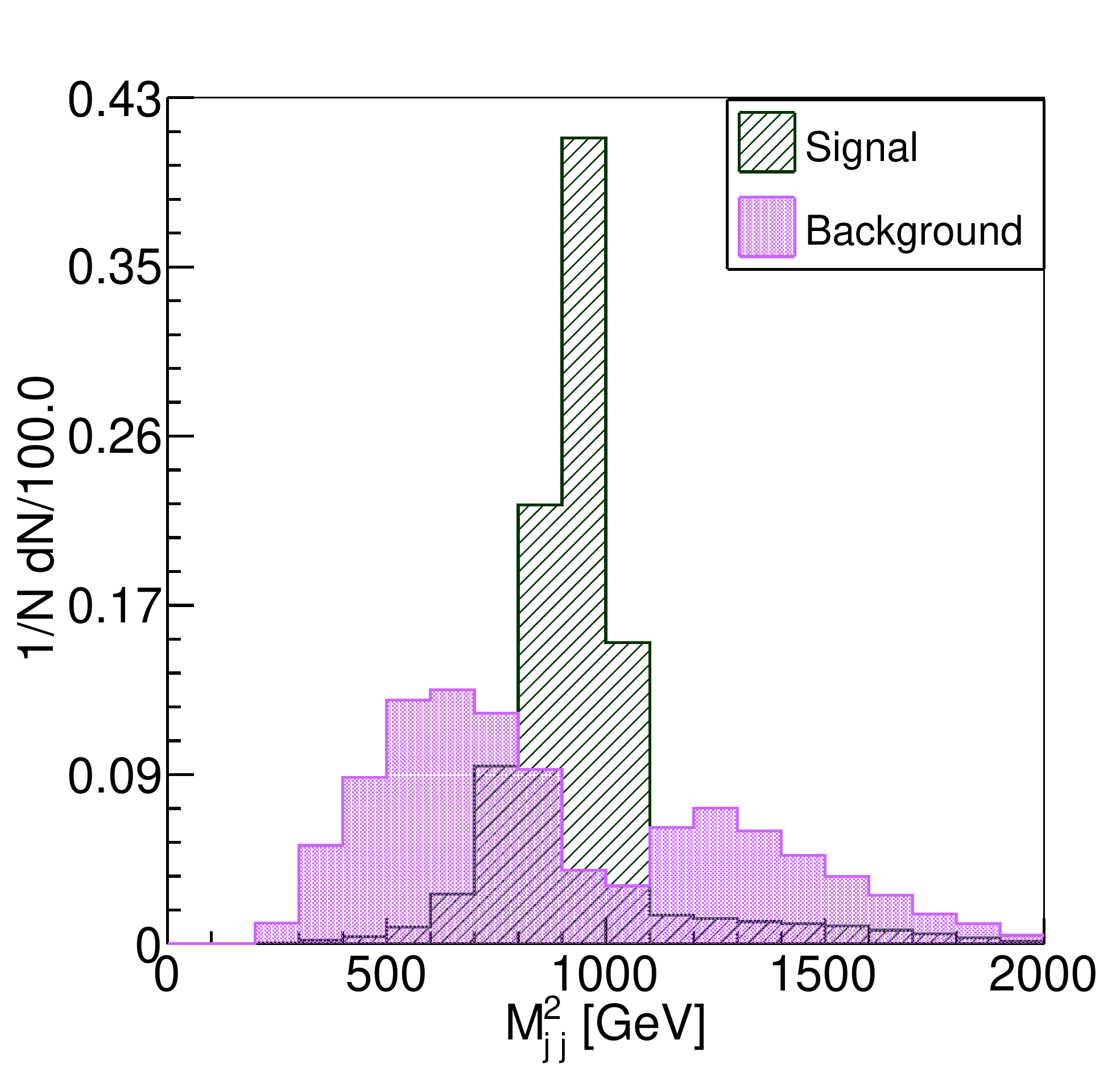}		
		\caption{\label{Fig:invmass_diss1} Normalized invariant mass distribution $M_{jj}^1$  (first pair) and $M_{jj}^2$  (second pair) for signal ($M_{H^{\pm\pm}}$ = 1 TeV) and background at 3 TeV $\mu^+$$\mu^-$  collider. }	
	\end{figure}
	
\end{widetext}

In Fig.\,\ref{Fig:jetinvmass_distribution}, we plot the invariant mass of $j_1$ ($M_{j1}$) and $j_4$ ($M_{j4}$) out of the leading 4 fat-jets. The distribution of $M_{j1}$ and $M_{j4}$ shows a clear peak around $M_{W^{\pm}}$. From this it is clear that the two prong W-jet submerged to form a single fat-jets. The small peak of $M_{j4}$ distribution signifies that the low $p_T$ jets are not pure W-jet. We also tried to reconstruct the invariant mass of the doubly-charged scalars ($H^{\pm\pm}$), out of the final state signal fat-jets. In Fig.\,\ref{Fig:invmass_diss1}, we plot the  two invariant mass reconstructed out of the final state fat-jets. $ M^1_{jj} $ and $  M^2_{jj} $ are the first and second pair of reconstructed invariant mass out of the four final fat-jets respectively. During the selection process of the second pair, we remove the those fat-jets which are already considered in construction the first pair, which takes care of the double counting. As we have shown the distribution for our benchmark point, $M_{H^{\pm\pm}} = 1000\, \text{GeV}$, we can see the invariant mass distributions clearly peaks around 1000 GeV for signal process.

To discriminate the signal from the background, we select events within 20 GeV mass window ($|M_{W^{\pm}} - M_{ji}|= \Delta M_{ji}  < 20\,\text{GeV}$) around $M_{W^{\pm}}$ for the leading three fat-jets. We also select events lying within a 100 GeV mass gap around the mass of $H^{\pm\pm}$ ($|M_{H^{\pm\pm}} - M^i_{jj}|= \Delta M^i_{jj}  < 100\, \text{GeV}$) to reduce the background further. We have summarized the results in cut flow Table.\,\ref{tab:cut_flow}.

	\begin{table*}[hbt!]
		\centering
		\begin{center}
			\begin{adjustbox}{width=1\textwidth}
				\small\addtolength{\tabcolsep}{3pt}
				\begin{tabular}{||c|c|c|c|c|c|c|c|c|c|c||}
				\hline 
				\multicolumn{11}{||c||}{ $\mu^+ \mu^-  \to H^{++} H^{--} \to W^{+} W^{+} W^{-} W^{-} \to N j_{\rm{fat}}$} \\ \hline \hline
				$M_{H^{\pm\pm}}$ (GeV)  & $\ge 4j$ & $p_T^{j1}$ $>450$  & $p_T^{j2}$ $>350$   &  $p_T^{j3}$ $>200$ & $p_T^{j4}$ $>100$ & $\Delta M_{j_1} < 20$ & $\Delta M_{j_2} < 20$ & $\Delta M_{j_3} < 20$ & $\Delta M^1_{jj}$ $<100$ & $\Delta M^2_{jj}$ $<100$ \\
				\hline
				1000\,[0.825] & 0.739 & 0.736 & 0.730 & 0.679 & 0.540 & 0.504 & 0.442 & 0.388 & 0.380 & 0.220\\
				1400\,[0.0922] & 0.064 & 0.0623 & 0.062 & 0.060 & 0.053 & 0.050 & 0.046 & 0.041 & 0.039 & 0.019\\
				\hline \hline
				\multicolumn{11}{||c||}{Backgrounds} \\ \hline \hline
				$\text{BG}$  & $\ge 4j$ & $p_T^{j1}$ $>450$  & $p_T^{j2}$ $>350$   &  $p_T^{j3}$ $>200$ & $p_T^{j4}$ $>100$ & $\Delta M_{j_1} < 20$ & $\Delta M_{j_2} < 20$ & $\Delta M_{j_3} < 20$ & $\Delta M^1_{jj}$ $<100$ & $\Delta M^2_{jj}$ $<100$ \\
				\hline
				$VV + 2j\,[4.680]$    & 1.527 & 1.390 & 1.256 & 0.351 & 0.156 & 0.066 & 0.028 & 0.0092 & 0.0053 & 0.0005 \\
				$V + 3j\,[5.450]$     & 3.613 & 3.319 & 3.077 & 1.815 & 0.907 & 0.284 & 0.088 & 0.0265 & 0.0151 & 0.002 \\
				$4j\,[6.677]$ 	     & 5.123 & 4.830 & 4.560 & 3.038 & 1.840 & 0.448 & 0.113 & 0.0288 & 0.0163 & 0.0019 \\
				$V + 2j\,[5.360]$     & 1.787 & 1.669 & 1.554 & 1.055 & 0.486 & 0.185 & 0.063 & 0.0195 & 0.0109 & 0.0013\\
				$VVV + 2j\,[0.049]$   & 0.048 & 0.034 & 0.027 & 0.021 & 0.015 & 0.0036 & 0.002 & 0.0017 & 0.001 & 0.00009 \\
				\hline 
				\end{tabular}
			\end{adjustbox}
			\caption{The cut-flow for the signal and backgrounds. For signal two benchmark $M_{H^{\pm\pm}}$ = 1000 GeV and $M_{H^{\pm\pm}}$ = 1400 GeV are considered. The values mentioned in square brackets are the production cross section of each processes. The cross sections are in $\textrm{fb}$.}
			\label{tab:cut_flow}
		\end{center} 	
	\end{table*}


\subsubsection{Cut-based Results}

\begin{figure}[h]
	\centering
	\includegraphics[scale=0.6]{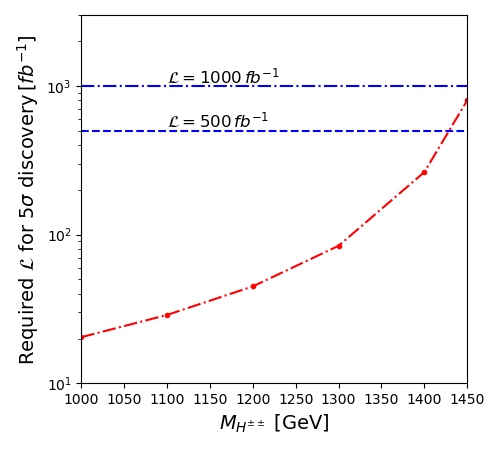}
	\caption{\label{Fig:required_lumi}Required luminosity for $5\sigma$ discovery of the doubly-charged scalars at 3 TeV $\mu^+$$\mu^-$  collider. }
\end{figure}

With the optimized signal and background events, that survive  the selection cuts, we analyse the signal sensitivity. With $s$ and $b$ being the signal cross section and  background cross section  after all the selection cuts, the statistical significance is defined as~\cite{Cowan:2010js,Cowan_talk:2012}
\begin{eqnarray}
	\sigma_{dis} &=& \Big[2\Big((s+b)ln\Big[1+\frac{s}{b}\Big]-s\Big)\Big]^{1/2}
	\label{Eq:discovery_no_error}
\end{eqnarray}
For a $5\sigma$ discovery we assume $\sigma_{dis}$ $\geq$ 5. The luminosity required for $5\sigma$ discovery for different benchmark mass of doubly charged scalar have been shown in Fig.\,\ref{Fig:required_lumi}.

\subsubsection{Multivariate Analysis (MVA) }
\begin{widetext}
	
	\begin{figure}[hbt!]
		\centering
		\includegraphics[scale=0.55]{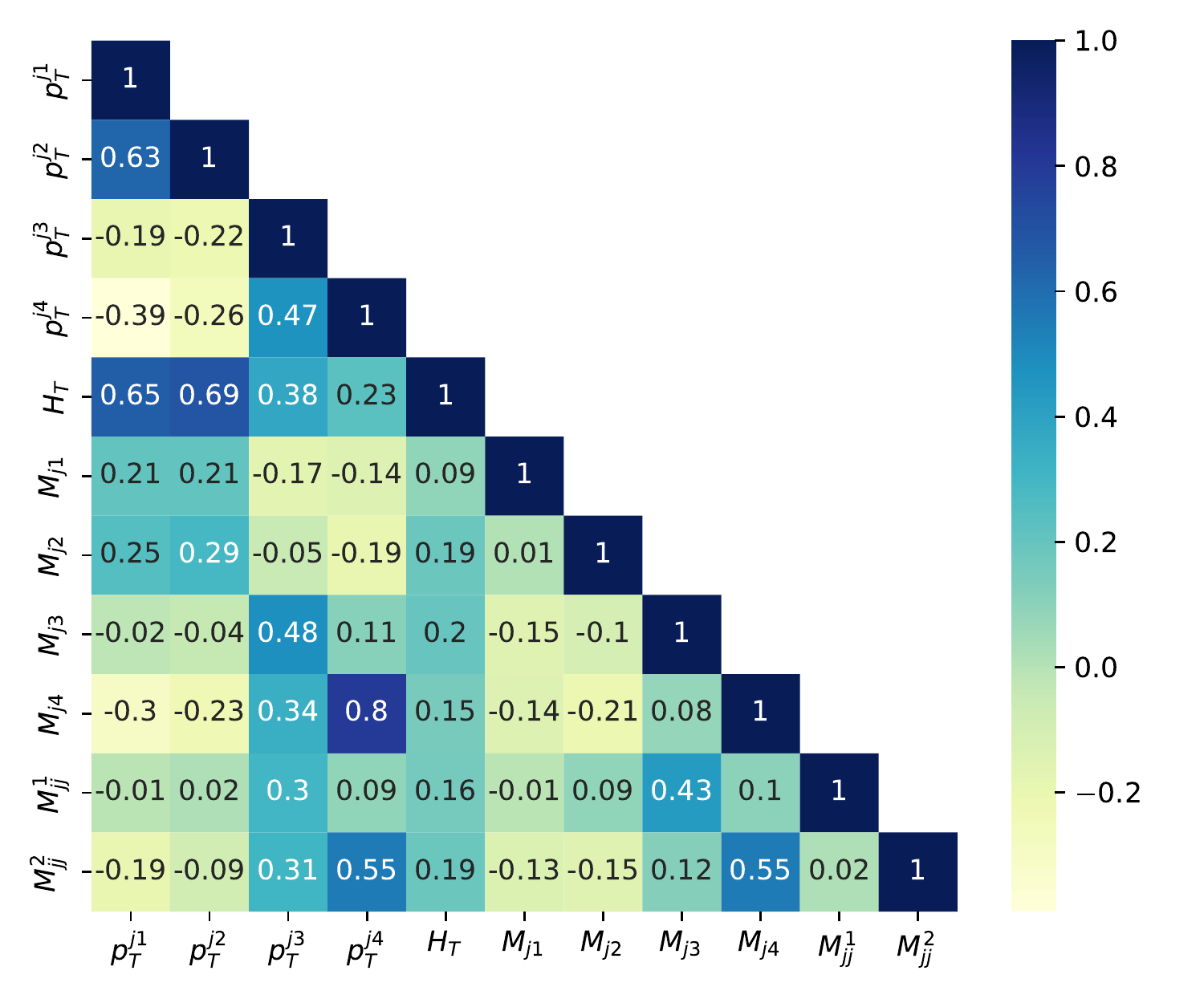}
		\includegraphics[scale=0.55]{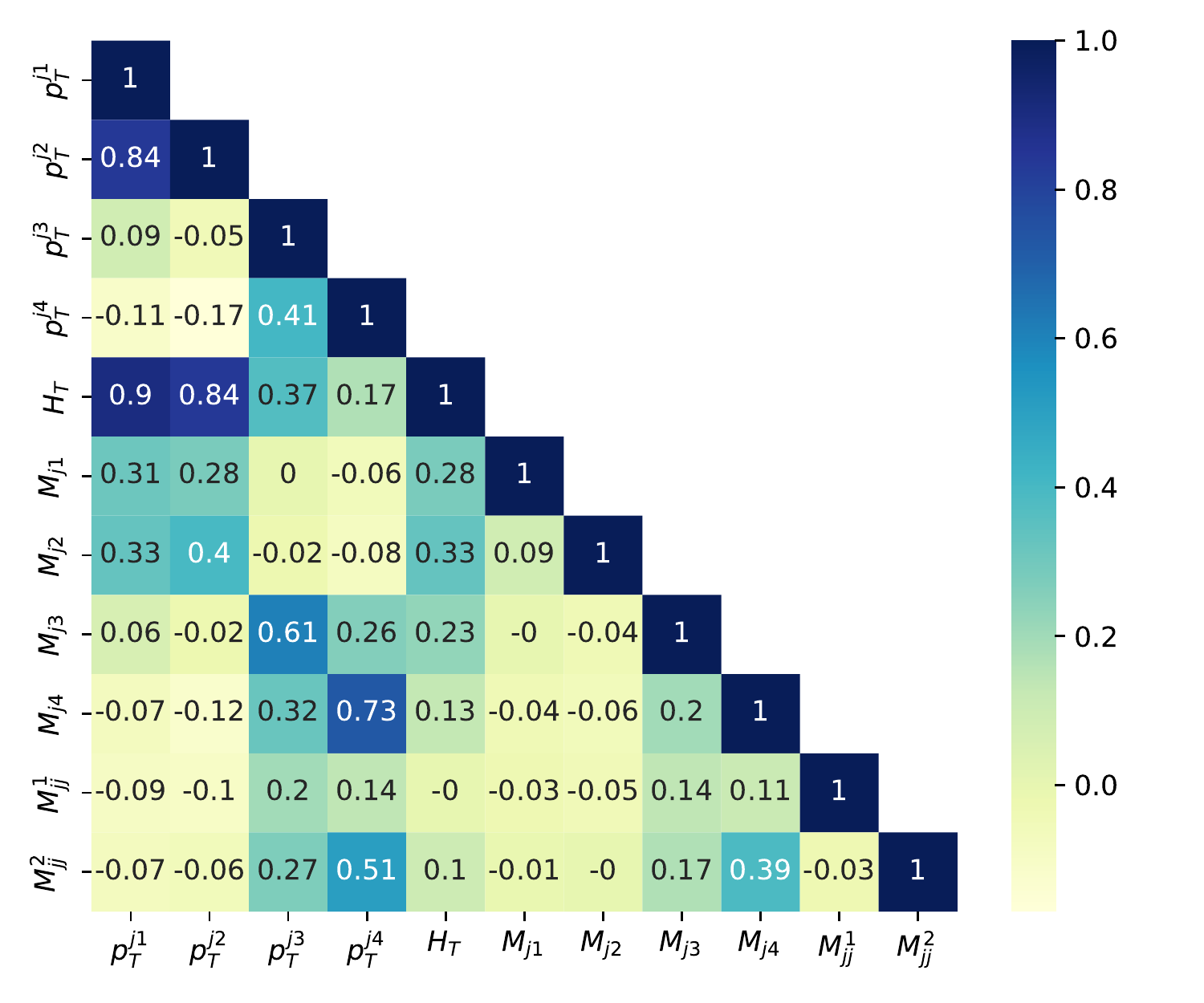}
		\caption{The linear correlation coefficients for signal (left) and background (right) between the different kinematic variables used in multivariate analysis.}
		\label{fig:correlation}
	\end{figure}
	
\end{widetext}

\begin{figure*}[htb!]
	\centering
	\includegraphics[height=9cm,width=15.5cm]{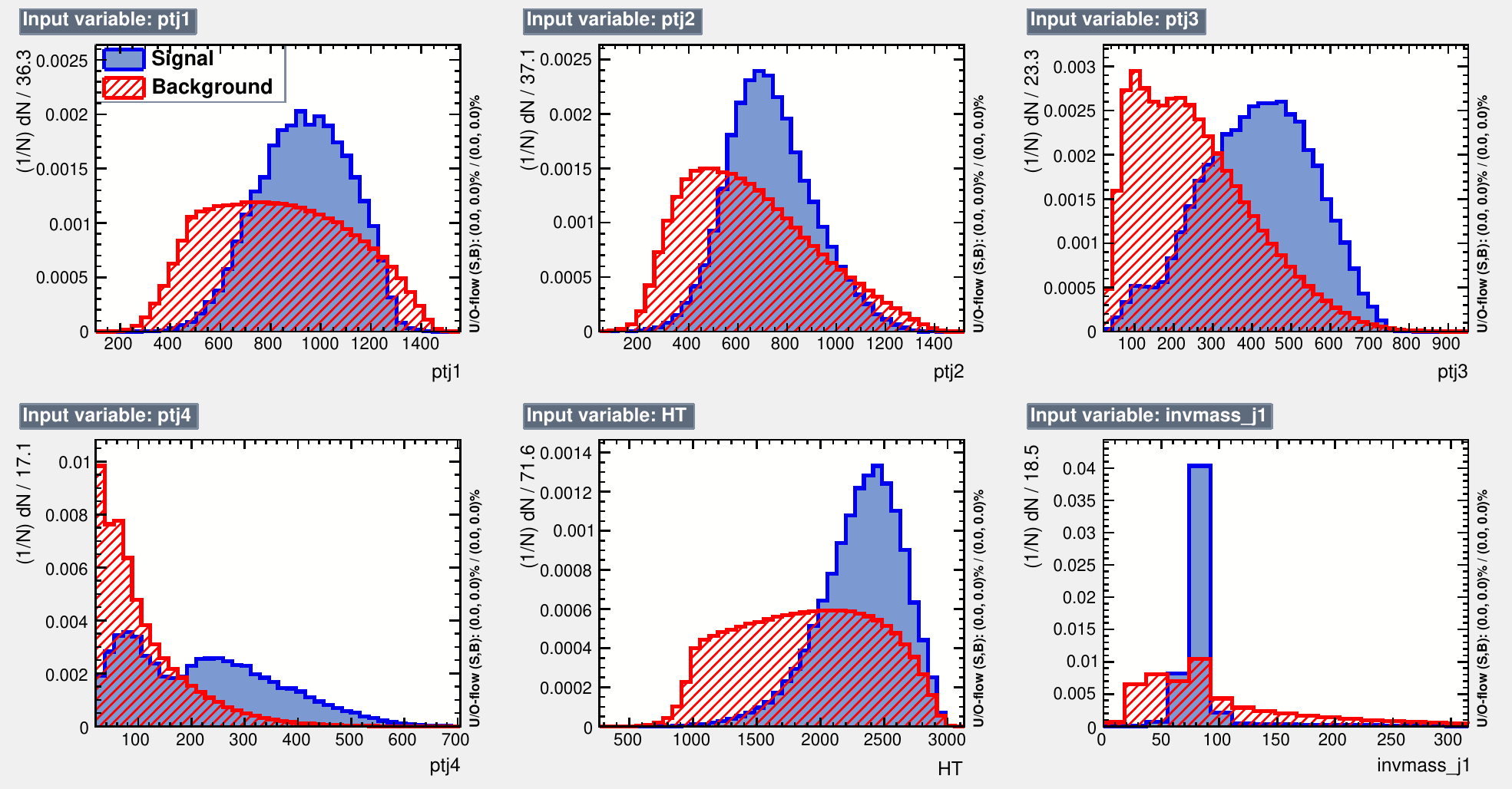}
	\includegraphics[height=9cm,width=15.5cm]{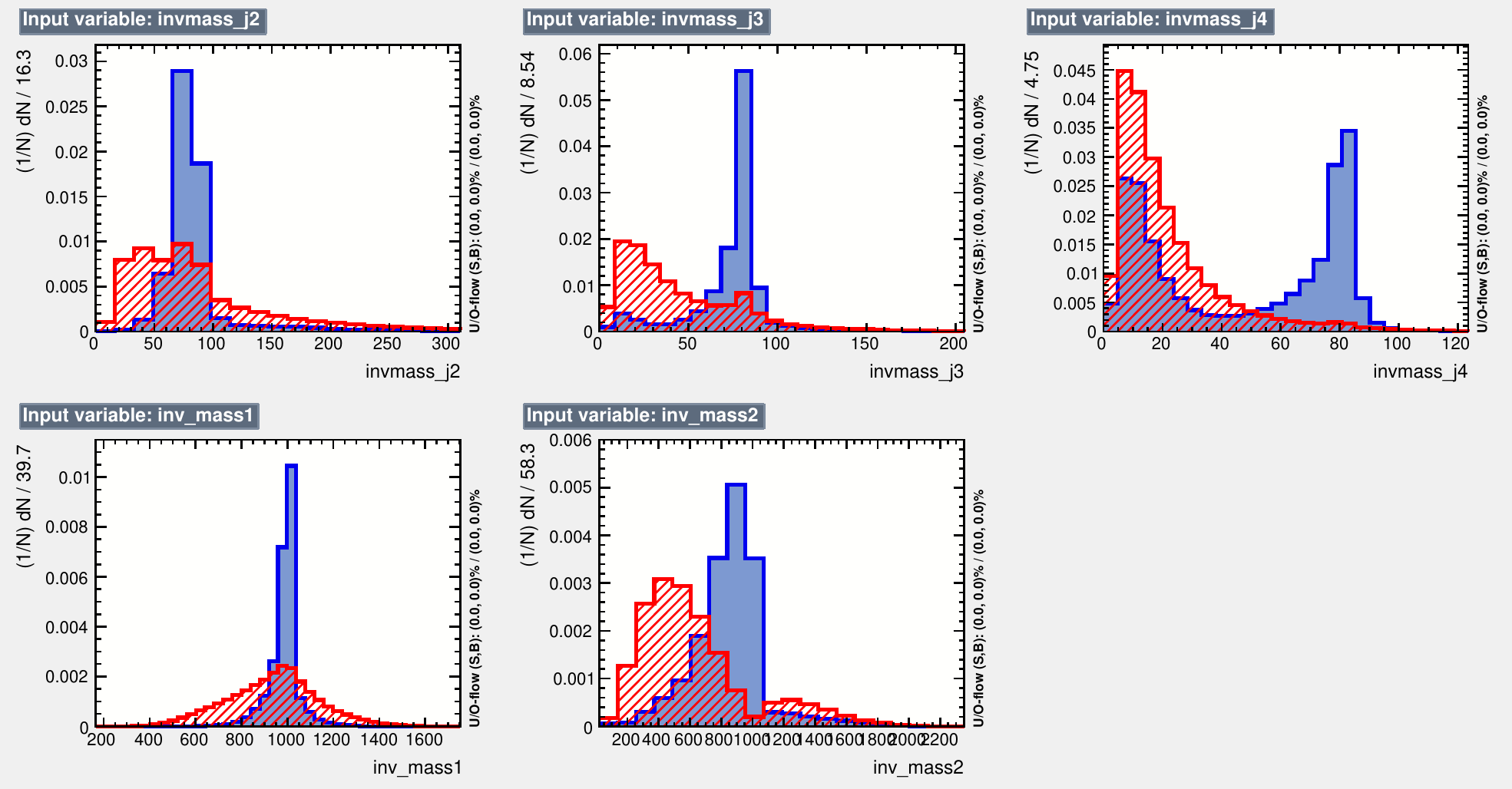}
	\caption{Distributions of different kinematic variables used for MVA for signal as well as for all back grounds. The normalised distributions for signal is given by the  solid blue line and the back ground distributions are given by solid red line.    }
	\label{Fig:BDT_variables}	
\end{figure*}

In this section we  discuss the results of the multivariate analysis (MVA) \cite{Hocker:2007ht} that we have performed. Fig.\,\ref{Fig:BDT_variables} shows the kinematic variables used for the MVA analysis assuming the variables are having a good discriminating power between the signal and background, and have low correlations among themselves. A detailed discussion about the variables used has already been presented in the text. The method-unspecific separation is a good measure of the discriminating power of any variable, for a given feature $y$ \cite{Hocker:2007ht}. This is defined as,
\begin{eqnarray}
	\left<S^{2}\right> = \frac{1}{2}\int dy
	\frac{(\hat{y}_S(y)-\hat{y}_B(y))^2}{\hat{y}_S(y)+\hat{y}_B(y)}
\end{eqnarray}
where $\hat{y}_S$ and $\hat{y}_B$ are the probability functions for the signal and background for the particular feature $y$. The quantity is equal to zero for similar signal and background distributions, and 1 for distributions with no overlaps. We have considered those variables having method-unspecific separation value greater than 1\%. In Table.\,\ref{tab:separting_variables}, we present the method-unspecific separation for the input variables. In Fig.\,\ref{fig:correlation}, we have shown the Pearson's linear correlation coefficients between the input variables, defined as,
\begin{eqnarray}
	\rho(x,y)=\frac{\left<xy\right> -\left<x\right>\left<y\right> }{\sigma_x \sigma_y}
\end{eqnarray}
where $\left<x\right>$ and $\sigma_x$ are the expectation value and standard deviation respectively for the dataset $x$.

In Table.\,\ref{tab:bdt_params}, we have summarised the relevant BDT hyperparameters. We have optimised the BDT  hyperparameters with \textit{adaptive boost} algorithm with a learning rate of 0.1. The so-called \textit{Gini index} has been used for the separation between the nodes in decision trees. In Table.\,\ref{tab:separting_variables}, we also present the method-specific ranking of the input variables, which shows the importance of the used variables in  separating signal from background. 

\begin{table}[htb!]
	\centering
	\begin{tabular}{lll}
		\toprule
		Feature & Method-unspecific & Method-specific \\ 
		& separation & ranking \\ 
		\midrule 
		$M_{j3}$    & 0.393 & 	    0.144\\
		$M_{j1}$    & 0.391 &	    0.142\\
		$M_{jj}^2$  & 0.384 &		0.102\\
		$M_{jj}^1$  & 0.336 &		0.095\\
		$M_{j2}$ 	& 0.289 & 		0.116\\
		$M_{j4}$    & 0.257 & 		0.096\\
		$p_T^{j3}$  & 0.241 & 		0.063\\
		$p_T^{j4}$  & 0.229 & 		0.045\\
		$H_T$      	& 0.218 &		0.066\\
		$p_T^{j1}$  & 0.141 &		0.057\\
		$p_T^{j2}$  & 0.137 & 		0.069\\ 
		\bottomrule
	\end{tabular} 
	\caption{\label{tab:separting_variables} Method unspesific and Method spesific relative importance of the different discriminating variables used in MVA.}
	
\end{table}

\begin{table}[htb!]
	\centering
	\begin{tabular}{ll}
		\toprule
		BDT hyperparameter & Optimised choice \\
		\midrule
		NTrees & $500$\tabularnewline
		MinNodeSize & $5.0$\%\tabularnewline
		MaxDepth & $5$\tabularnewline
		BoostType & AdaBoost\tabularnewline
		AdaBoostBeta & $0.1$\tabularnewline
		UseBaggedBoost & True\tabularnewline
		BaggedSampleFraction & $0.5$\tabularnewline
		SeparationType & GiniIndex\tabularnewline
		nCuts & 20\tabularnewline 
		\bottomrule
	\end{tabular} 
\caption{Summary of optimised BDT hyperparameters.}
\label{tab:bdt_params}
\end{table}

\begin{widetext}
	
\begin{figure*}[htb!]
	\centering
	\includegraphics[width=0.48\textwidth]{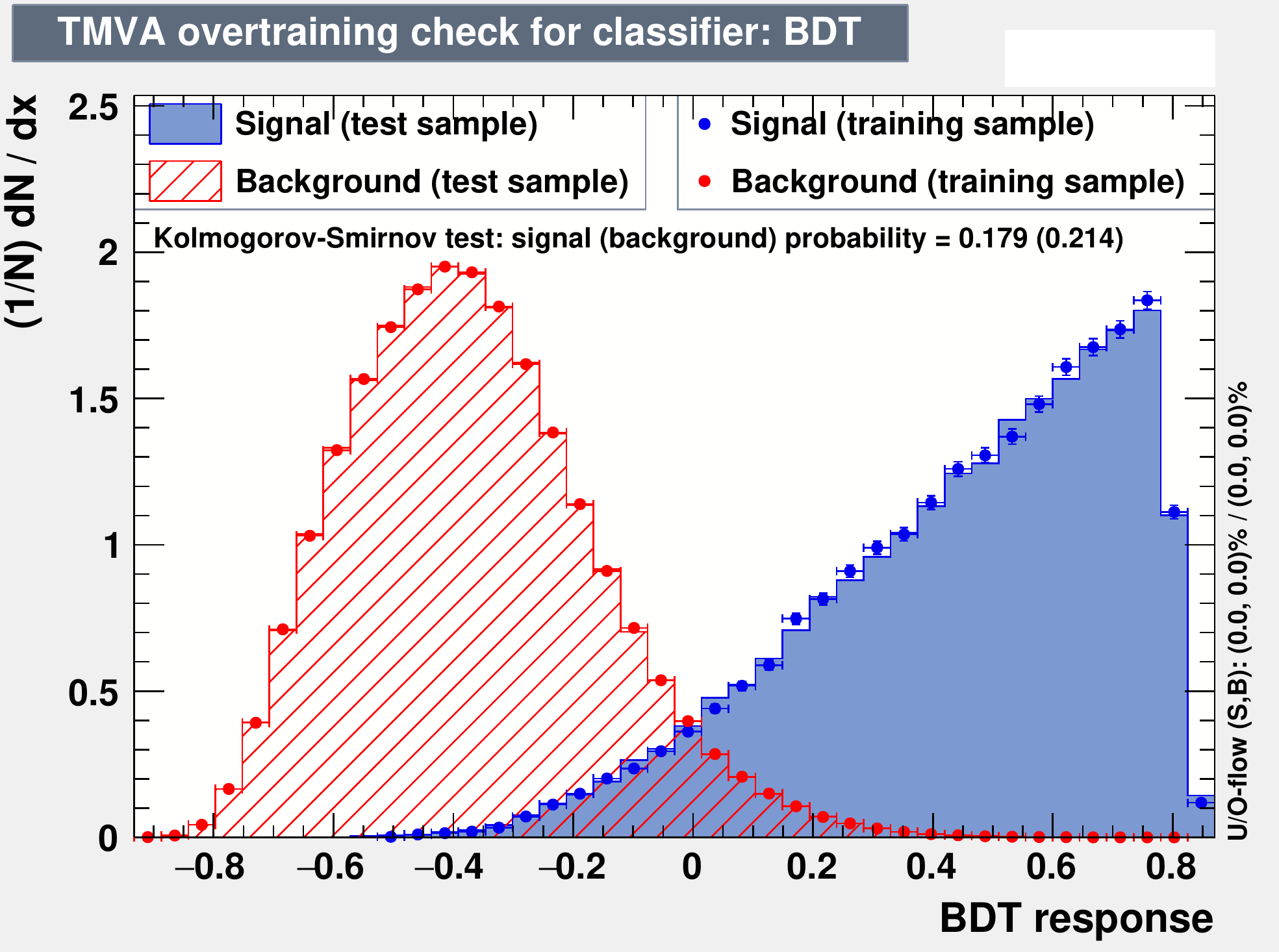}
	\includegraphics[width=0.48\textwidth]{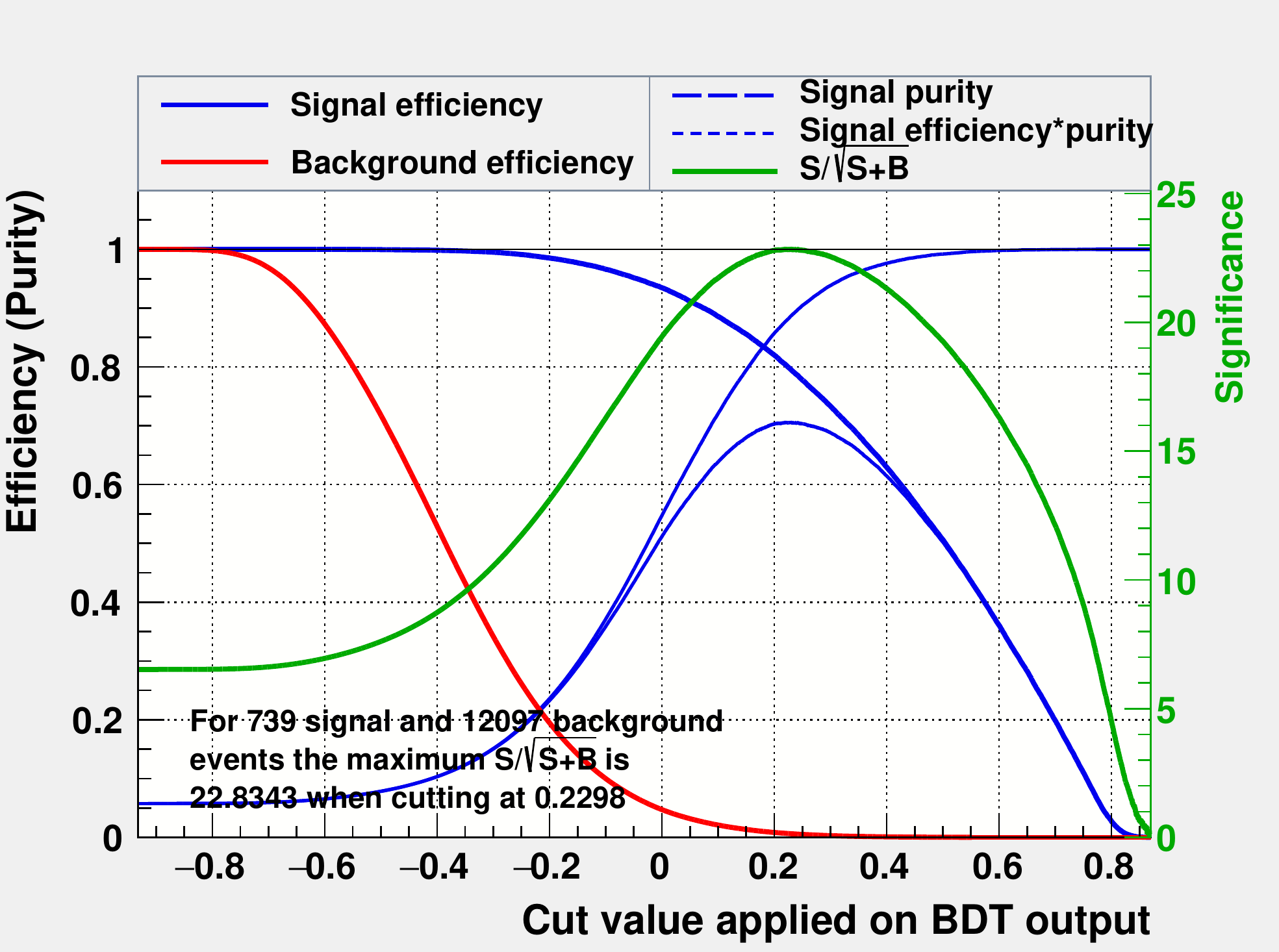}
	\caption{(Left): Normalised distribution of the BDT response for both signal(blue) and background(red) classes (both training and testing samples). (Right): Signal(blue) and background(red) efficiencies along with the statistical significance(green) of the signal as function of applied BDT cut value. }
	\label{fig:BDT_res_cuteff}	
\end{figure*}

\end{widetext}

We present the BDT response and cut efficiency at a benchmark mass value, $M_{H^{\pm\pm}}$ = 1000 GeV, in Fig.\,\ref{fig:BDT_res_cuteff}. We have used the optimised hyperparameters at different benchmark mass to obtain the statistical significance, 
\begin{eqnarray}
	\mathcal{Z} = \frac{N_S}{\sqrt{N_S + N_B}} 
\end{eqnarray}
where $N_S$ and $N_B$ are the number of signal and background events after the optimal cut is applied on the BDT response. In Table.\,\ref{tab:BDT_significance}, we present the obtained statistical significance $\mathcal{Z}$ for different benchmark doubly-charged scalar mass $M_{H^{\pm\pm}}$.

\begin{table}[htb!]
	\centering %
	\begin{tabular}{cccc}
		\toprule %
		\toprule
		Mass[GeV] & Significance\\ \toprule %
		1000    & 22.834\\
		1100    & 19.303\\
		1200    & 15.256\\
		1300    & 10.959\\
		1400    & 6.259\\
		1450    & 3.640\\ \bottomrule%
	\end{tabular}
	\caption{Statistical significance, $\mathcal{Z}$, for different doubly-charged scalar mass obtained from the multivariate analysis for muon collider with center of mass $\sqrt{s}$ = 3 TeV and integrated luminosity $\mathcal{L}$ = 1000 $\textrm{fb}^{-1}$.}
		\label{tab:BDT_significance}
\end{table}

\section{\label{sec:4}Summary}
	We have discussed the discovery prospects of the doubly-charged scalar, $H^{\pm\pm}$, present in Type-II seesaw model in $\mu^+\mu^-$ collider. We mainly focus our attention to that part of the parameter space where the produced doubly-charged scalar mainly decays into $W^\pm W^\pm$ final state and subsequently we consider the hadronic final states from the $W$ decays. All hadronic final states are not favourable channel to probe at the LHC   due to the presence of the towering QCD backgrounds. Hence this region is more favorable for linear colliders, \textit{viz} $e^+e^-$ and $\mu^+\mu^-$ collider. Among these two, the $\mu^+\mu^-$ collider has a low energy loss due to synchrotron radiation. Firstly, we have performed the trivial cut-based analysis and predict the statistical significance of the muon collider at $\sqrt{s} = 3$ TeV. Secondly, we have also performed a multivariate analysis 	and compared both the results. From the cut-based analysis we have concluded that at 1000 $\textrm{fb}^{-1}$ luminosity up to 1450 GeV massive doubly-charged scalar can be discovered  with $5\sigma$ confidence level, see Fig.\,\ref{Fig:required_lumi}. The result from the multivariate analysis for different $M_H^{\pm\pm}$ value has been given in Table.\,\ref{tab:BDT_significance}, from which it is evident that, apart from a higher mass of 1.450 TeV, BDT analysis offers a larger statistical significance, greater than $5\sigma$. 
	
\section{\label{sec:5}Acknowledgments}
The authors acknowledge SAMKHYA: high Performance Computing  Facility provided by Institute of physics, for the performed simulations. The authors acknowledge the support from the Indo-French Centre for the Promotion of Advanced Research (Grant no: 6304-2). SPM thank Saiyad Ashanujjaman, Rojalin Padhan and Agnivo Sarkar for the useful discussions.

\bibliography{main.bib}
\end{document}